\documentclass[gmd, manuscript]{copernicus}

\newcommand{\Sympy}{{\em SymPy} }

\newcommand{\Python}{{\em Python} }
\newcommand{\SciPy}{{\em SciPy} }
\newcommand{\Matlab}{Matlab\textsuperscript{\tiny{\textregistered}}}

\nolinenumbers

\begin{document}

\title{Devito (v3.1.0): an embedded domain-specific language for finite differences and geophysical exploration}


\Author[1]{Mathias}{Louboutin}
\Author[2]{Michael}{Lange}
\Author[3]{Fabio}{Luporini}
\Author[3]{Navjot}{Kukreja}
\Author[1]{Philipp A.}{Witte}
\Author[1]{Felix J.}{Herrmann}
\Author[3]{Paulius}{Velesko}
\Author[3]{Gerard J.}{Gorman}

\affil[1]{School of Computational Science and Engineering, Georgia Institute of Technology, Atlanta, USA}
\affil[2]{ECMWF, UK}
\affil[3]{Earth Science and Engineering, Imperial College London, London, UK}


\runningtitle{Devito: an embedded domain-specific language for finite differences and geophysical exploration}

\runningauthor{Mathias Louboutin}

\correspondence{Mathias Louboutin (mlouboutin3@gatech.edu)}

\received{}
\pubdiscuss{} 
\revised{}
\accepted{}
\published{}


\firstpage{1}

\maketitle

\begin{abstract}

We introduce Devito, a new domain-specific language for implementing
high-performance finite difference partial differential equation solvers.
The motivating application is exploration seismology where methods such as
Full-Waveform Inversion and Reverse-Time Migration are used to
invert terabytes of seismic data to create images of the earth's
subsurface. Even using modern supercomputers, it can take weeks to process a
single seismic survey and create a useful subsurface image. The
computational cost is dominated by the numerical solution of wave equations and
their corresponding adjoints. Therefore, a great deal of effort is invested in
aggressively optimizing the performance of these wave-equation propagators
for different computer architectures. Additionally, the actual set of
partial differential equations being solved and their numerical
discretization is under constant innovation as increasingly realistic
representations of the physics are developed, further ratcheting up the
cost of practical solvers. By embedding a domain-specific language within
Python and making heavy use of SymPy, a symbolic mathematics library, we
make it possible to develop finite difference simulators quickly using a
syntax that strongly resembles the mathematics. The Devito compiler reads
this code and applies a wide range of analysis to generate highly optimized
and parallel code. This approach can reduce the development time of a
verified and optimized solver from months to days.

\end{abstract}

\introduction  
Large-scale inversion problems in exploration seismology constitute some of the
most computationally demanding problems in industrial and academic research.
Developing computationally efficient solutions for applications such as seismic
inversion requires expertise ranging from theoretical and numerical methods
for partial differential equation (PDE) constrained optimization to low-level
performance optimization of PDE solvers. Progress in this area is often
limited by the complexity and cost of developing bespoke wave propagators (and
their discrete adjoints) for each new inversion algorithm or formulation of wave
physics.  Traditional software engineering approaches often lead developers to
make critical choices regarding the numerical discretization before manual
performance optimization for a specific target architecture and making it ready
for production. This workflow of bringing new algorithms into production, or
even to a stage that they can be evaluated on realistic datasets can take many
person-months or even person-years. Furthermore, it leads to mathematical
software that is not easily ported, maintained or extended. In contrast, the
use of high-level abstractions and symbolic reasoning provided by
domain-specific languages (DSL) can significantly reduce the time it takes to
implement and verify individual operators for use in inversion frameworks, as
has already been shown for the finite element method \citep{Fenics, Firedrake,
DolfinAdjoint}.

State-of-the-art seismic imaging is primarily based upon explicit finite
difference schemes due to their relative simplicity and ease of
implementation~\citep{Virieux86, symes2015acoustic, Weiss2013}. When
considering how to design a DSL for explicit finite difference schemes, it is
useful to recognize the algorithm as being primarily a sub-class of stencil
algorithms or polyhedral computation \citep{StencilDSL-1, Andreolli2015, YASK}.
However, stencil compilers lack two significant features required to
develop a DSL for finite differences: symbolic computational support required
to express finite difference discretizations at a high level, enabling these
expressions to be composed and manipulated algorithmically; support for
algorithms that are not stencil-like, such as source and receiver terms that are
both sparse and unaligned with the finite difference grid. Therefore, the design
aims behind the Devito DSL can be summarized as:
\begin{itemize}
\item create a high-level mathematical abstraction for programming finite
differences to enable composability and algorithmic optimization,
\item insofar as possible use existing compiler technologies
to optimize the affine loop nests of the computation, which account for most
of the computational cost,
\item develop specific extensions for other parts of the
computation that are non-affine (e.g., source and receiver terms).
\end{itemize}

The first of these aims is primarily accomplished by embedding the DSL in \Python
and leveraging the symbolic mathematics package Sympy~\citep{Sympy}.  From this
starting point, an abstract syntax tree is generated and standard compiler
algorithms can be employed to either generate optimized and parallel C code
or to write code for a stencil DSL - which itself will be passed to the next
compiler in the chain. The fact that this can be all performed just-in-time
(JIT) means that a combination of static and dynamic analysis can be used to
generate optimized code. However, in some circumstances, one might also choose
to compile offline.

The use of symbolic manipulation, code generation, and just-in-time compilation
allows the definition of individual wave propagators for inversion in only a
few lines of \Python code, while aspects such as varying the problem
discretization become as simple as changing a single parameter in the problem
specification, for example, changing the order of the spatial discretization
\citep{louboutin2016ppf}. This article explains {\it what} can be accomplished
with Devito, showing how to express real-life wave propagators as well as their
integration within larger workflows typical of seismic exploration, such as the
popular Full-Waveform Inversion (FWI) and Reverse-Time Migration (RTM) methods.
The Devito compiler, and in particular {\it how} the user-provided SymPy
equations are translated into high-performance C, are also briefly
summarized, although for a complete description the interested reader should
refer to~\citet{devito-compiler}.

The remainder of this paper is structured as follows: first, we provide a brief
history of optimizing compilers, DSL and existing wave equation seismic
frameworks. Next, we highlight the core features of Devito and describe the
implementation of the featured wave equation operators in
Sec.~\ref{sec:devito}. We outline the seismic inversion theory in
Sec.~\ref{sec:seismic}. Code verification and analysis of accuracy in
Sec.~\ref{sec:tests} is followed by a discussion of the propagators
computational performance in Sec.~\ref{sec:performances}. We conclude by
presenting a set of realistic examples such as seismic inversion and
computational fluid dynamics and a discussion of future work.

\section{Background} \label{Background}
Improving the runtime performance of a critical piece of code on a particular
computing platform is a non-trivial task that has received significant
attention throughout the history of computing. The desire to automate the
performance optimization process itself across a range of target architectures
is not new either, although it is often met with skepticism. Even the very
first compiler, A0 \citep{hopperA0}, was received with resistance, as best
summarized in the following quote: \emph{``Dr. Hopper believes,..., that the
result of a compiling technique should be a routine just as efficient as a hand
tailored routine. Some others do not completely agree with this. They feel the
machine-made routine can approach hand tailored coding, but they believe there
are "tricks of the trade" that apply to various special cases that a computer
cannot be expected to utilize."}~\citep{jones1954survey}. Given the challenges
of porting optimized codes to a wide range of rapidly evolving computer
architectures, it seems natural to raise again the layer of abstraction
and use compiler techniques to replace much of the manual labor.

Community acceptance of these new ``automatic coding systems'' began when
concerns about the performance of the generated code were addressed by the
first ``optimizing compiler'', FORTRAN, released in 1957 -- which not only
translated code from one language to another but also ensured that the final
code performed at least as good as a hand-written low-level
code~\citep{HistoryOfFortran}. Since then, as program and hardware complexity
rose, the same problem has been solved over and over again, each time by the
introduction of higher levels of abstractions. The first high-level languages
and compilers were targeted at solving a large variety of problems and hence
were restricted in the kind of optimizations they could leverage. As these
generic languages became common-place and the need for further improvement in
performance was felt, restricted languages focusing on smaller problem domains
were developed that could leverage more ``tricks of the trade'' to optimize
performance. This led to the proliferation of DSLs for broad mathematical
domains or sub-domains, such as APL~\citep{Iverson1962}, Mathematica, \Matlab
or R.

In addition to these relatively general mathematical languages, more
specialized frameworks targeting the automated solution of PDEs have long been
of interest~\citep{PDEL,deqsol,alpal,ctadel}. More recent examples not only aim
to encapsulate the high-level notation of PDEs, but are often centered around a
particular numerical method. Two prominent contemporary projects based on the
finite element method (FEM), FEniCS~\citep{Fenics} and
Firedrake~\citep{Firedrake}, both implement a common DSL,
UFL~\citep{Alnaes2014}, that allows the expression of variational problems in
weak form. Multiple DSLs to express stencil-like algorithms have also
emerged over time, including \citep{StencilDSL-1,Autotuning-1,PATUS,
unat2011mint, koster2014platform, HiPACC, osuna2014stella, tang2011pochoir,
pluto, YASK}. Other stencil DSLs have been developed with the objective of
solving PDEs using finite differences \citep{STARGATES, simflowny, JacobsSBLI}. However, in all
cases their use in seismic imaging problems (or even more broadly in science
and engineering) has been limited by a number of factors other than technology
inertia. Firstly, they only raise the abstraction to the level of
polyhedral-like (affine) loops. As they do not generally use a symbolic
mathematics engine to write the mathematical expressions at a high-level,
developers must still write potentially complex numerical kernels in the target
low-level programming language. For complex formulations, this process can be
time-consuming and error prone, as hand-tuned codes for wave propagators can
reach thousands of lines of code. Secondly, most DSLs rarely offer enough
flexibility for extension beyond their original scope
(e.g. sparse operators for source terms and interpolation) making it difficult
to work the DSL into a more complex science/engineering workflow. Finally,
since finite difference wave propagators only form part of the over-arching PDE
constrained (wave equation) optimization problem, composability with external
packages, such as the \SciPy optimization toolbox, is a key requirement
that is often ignored by self-contained standalone DSLs. The use of a fully
embedded \Python DSL, on the other hand, allows users to leverage a variety of
higher-level optimization techniques through a rich variety of software
packages provided by the scientific \Python ecosystem.

Moreover, several computational frameworks for seismic imaging exist, although
they provide varying degrees of abstraction and are typically limited to a
single representation of the wave equation. IWAVE~\citep{GPR:GPR977,
symes2015iwave, sun2010iwave, symes2015acoustic}, although not a DSL,
provides a high-level of abstraction and a mathematical framework to abstract the
algebra related to the wave-equation and its solution. IWAVE provides a
rigorous mathematical abstraction for linear operations and vector
representations including Hilbert space abstraction for norms and distances.
However, its C++ implementation limits the extensibility of the framework to
new wave-equations. Other software frameworks, such as
Madagascar~\citep{Madagascar}, offer a broad range of applications. Madagascar
is based on a set of subroutines for each individual problem and offers
modelling and imaging operators for multiple wave-equations. However, the lack
of high-level abstraction restricts its flexibility and interfacing with high-level external software (i.e \Python, {\em Java}). The subroutines are also
mostly written in C/Fortran and limit the architecture portability.

\section{Symbolic definition of finite difference stencils with Devito}
\label{sec:devito}

In general, the majority of the computational workload in wave-equation based
seismic inversion algorithms arises from computing solutions to discrete wave
equations and their adjoints.  There are a wide range of mathematical models
used in seismic imaging that approximate the physics to varying degrees of
fidelity. To improve development and innovation time, including code
verification, we describe the use of the symbolic finite difference framework
Devito to create a set of explicit matrix-free operators of arbitrary spatial
discretization order. These operators are derived for example from the acoustic
wave equation
\begin{equation}\label{eq:acoustic}
\begin{aligned}
m(x) \frac{\partial^2{u(t, x)}}{\partial{t^2}} - \Delta u(t, x)= q(t, x),
\end{aligned}
\end{equation}
where $m(x)=\frac{1}{c(x)^2}$ is the squared slowness with $c(x)$ the spatially
dependent speed of sound, symbol $\Delta u(t, x)$ denotes the Laplacian of the
wavefield $u(t, x)$ and $q(t, x)$ is a source usually located at a single
location $x_s$ in space ($q(t, x) = f(t)\delta(x_s)$). This formulation will be
used as running example throughout the section.

\subsection{Code generation - an overview}

\begin{figure}\centering
  \includegraphics[width=.95\textwidth]{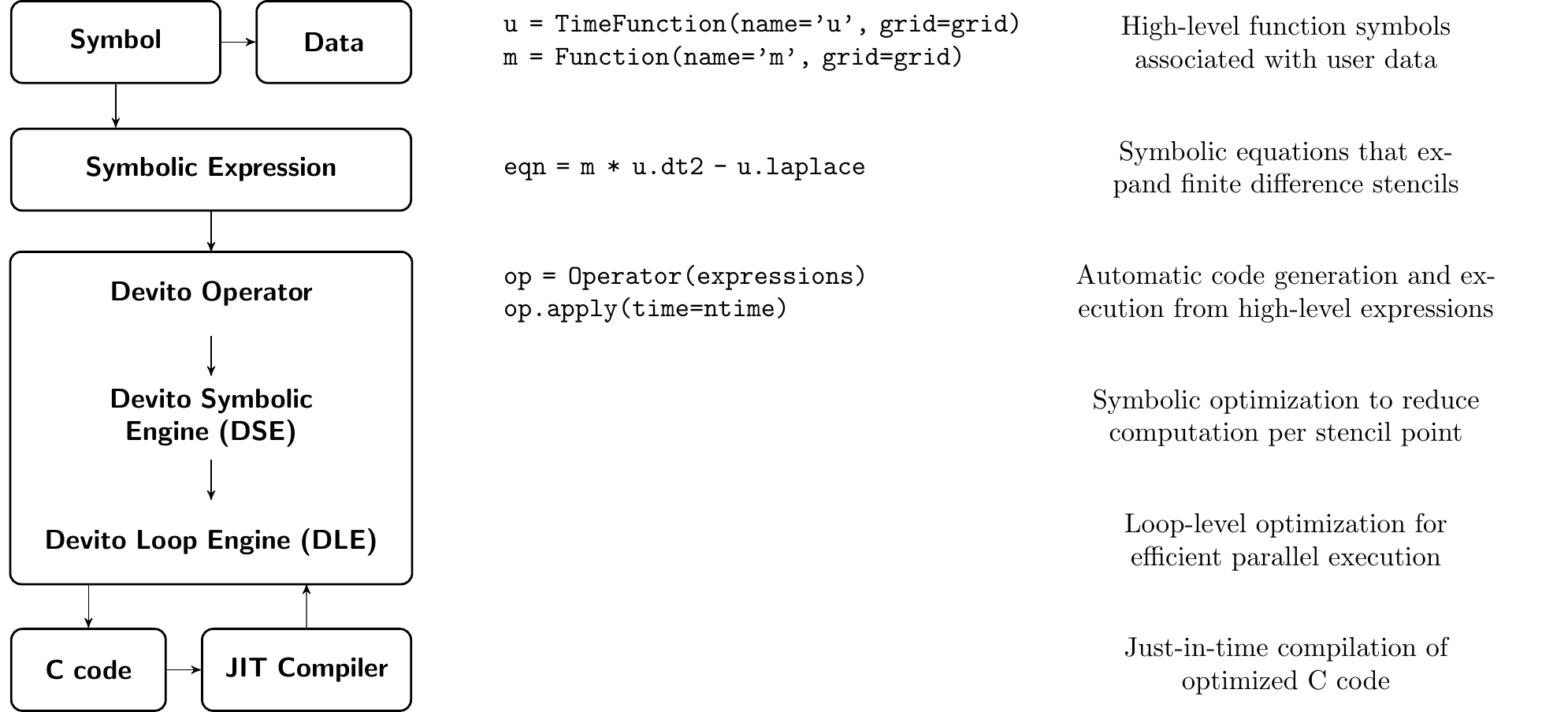}
  \caption{
    Overview of the Devito architecture and the associated example
    workflow. Devito's top-level API allows users to generate symbolic
    stencil expressions using data-carrying function objects that can
    be used for symbolic expressions via \Sympy. From this
    high-level definition, an operator then generates, compiles and
    executes high-performance C code.
  }
  \label{fig:architecture}
\end{figure}

Devito aims to combine performance benefits of dedicated stencil
frameworks~\citep{pluto, tang2011pochoir, StencilDSL-1, YASK} with the
expressiveness of symbolic PDE-solving DSLs~\citep{Fenics, Firedrake} through
automated code generation and optimization from high-level symbolic expressions
of the mathematics. Thus, the primary design objectives of the Devito DSL are
to allow users to define explicit finite difference operators for
(time-dependent) PDEs in a concise symbolic manner and provide an API that is
flexible enough to fully support realistic scientific use cases. To this end,
Devito offers a set of symbolic classes that are fully compatible with the
general-purpose symbolic algebra package \Sympy that enables users to derive
discretized stencil expressions in symbolic form. As we show in
Fig.~\ref{fig:architecture}, the primary symbols in such expressions are
associated with user data that carry domain-specific meta-data information to be
used by the compiler engine (e.g. dimensions, data type, grid). The discretized
expressions form an abstract operator definition that Devito uses to generate
low-level C code (C99) and OpenMP at runtime. The encapsulating
\texttt{Operator} object can be used to execute the generated code from within
the \Python interpreter making Devito natively compatible with the wide range of
tools available in the scientific \Python software stack. We manage memory using our own allocators (e.g. to enforce alignment and NUMA optimizations) and therefore we also take control over freeing memory. We wrap everything with the NumPy array API to ensure interoperability with other modules that use NumPy.

A Devito \texttt{Operator} takes as input a collection of symbolic expressions
and progressively lowers the symbolic representation to semantically equivalent
C code. The code generation process consists of a sequence of compiler passes
during which multiple automated performance-optimization techniques are
employed. These can be broadly categorised into two classes and are performed
by distinct sub-packages:

\begin{itemize}
\item \textbf{Devito Symbolic Engine (DSE):} Symbolic optimization
  techniques, such as Common Sub-expression Elimination (CSE),
  factorization and loop-invariant code motion are utilized to reduce
  the number of floating point operations (flops) performed within the
  computational kernel~\citep{Luporini2015}. These optimization techniques are inspired by \Sympy but are custom implemented in Devito and do not rely on \Sympy implementation of CSE for example.

\item \textbf{Devito Loop Engine (DLE):} Well-known loop optimization
  techniques, such as explicit vectorization, thread-level
  parallelization and loop blocking with auto-tuned block sizes are
  employed to increase the cache utilization and thus memory bandwidth
  utilization of the kernels.
\end{itemize}

A complete description of the compilation pipeline is provided
in~\citet{devito-compiler}.

\subsection{Discrete function symbols}

The primary user-facing API of Devito allows the definition of complex stencil
operators from a concise mathematical notation. For this purpose, Devito relies
strongly on \Sympy (Devito 3.1.0 depends upon SymPy 1.1 and all dependency versions are specified in Devito's requirements file).
Devito provides two symbolic object types that mimic \Sympy
symbols, enabling the construction of stencil expressions in symbolic form:

\begin{itemize}

\item \textbf{Function:} The primary class of symbols provided by Devito
  		behaves like \texttt{sympy.Function} objects, allowing symbolic
        differentiation via finite difference discretization and general
        symbolic manipulation through \Sympy utilities. Symbolic function
        objects encapsulate state variables (parameters and solution of the
        PDE) in the operator definition and associated user data (function
        value) with the represented symbol. The meta-data, such as grid
        information and numerical type, which provide domain-specific information
        to the Devito compiler are also carried by the \texttt{sympy.Function}
        object.

\item \textbf{Dimension:} Each \texttt{sympy.Function} object defines an
    iteration space for stencil operations through a set of \texttt{Dimension}
        objects that are used to define and generate the corresponding loop
        structure from the symbolic expressions.

\end{itemize}

\begin{figure}[h]
\begin{center}
\raggedright
  \includegraphics[]{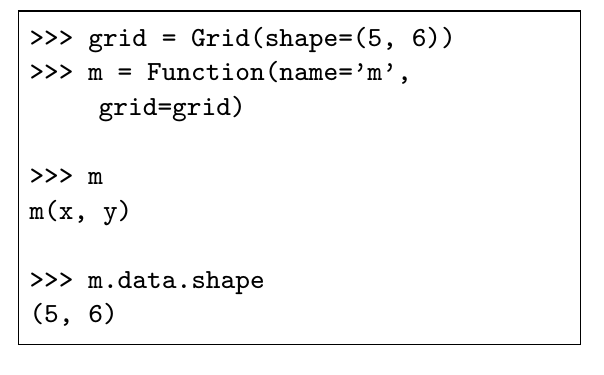}
    \caption{Defining a Devito {\tt Function} on a {\tt Grid}.}
    \label{lst:grid_function}
\end{center}
\end{figure}

In addition to \texttt{sympy.Function} and \texttt{Dimension} symbols, Devito
supplies the construct \texttt{Grid}, which encapsulates the definition of
the computational domain and defines the discrete shape (number of grid points,
grid spacing) of the function data. The number of spatial dimensions is
hereby derived from the shape of the \texttt{Grid} object and inherited by all
\texttt{Function} objects, allowing the same symbolic operator definitions to
be used for two and three-dimensional problem definitions. As an example, a
two-dimensional discrete representation of the square slowness of an acoustic
wave $\vec{m}[x, y]$ inside a 5 by 6 grid points domain can be created as a
symbolic function object as illustrated in Fig.~\ref{lst:grid_function}.

It is important to note here that $\vec{m}[x, y]$ is constant in time, while
the discrete wavefield $\vec{u}[t, x, y]$ is time-dependent. Since time is
often used as the stepping dimension for buffered stencil operators, Devito
provides an additional function type \texttt{TimeFunction}, which automatically
adds a special \texttt{TimeDimension} object to the list of dimensions.
\texttt{ TimeFunction} objects derive from \texttt{Function} with an extra time dimension
and inherit all the symbolic properties. The creation of a \texttt{TimeFunction} requires the same
parameters as a \texttt{Function}, with an extra optional \texttt{time\_order} property that defines
the discretization order for the time dimension and an integer \texttt{save} parameter that defines the
size of the time axis when the full time history of the field is stored in memory. In the case of a buffered time dimension \texttt{save} is equal to \texttt{None} and
the size of the buffered dimension is automatically inferred from the \texttt{time\_order} value.
As an example, we can create an equivalent symbolic representation
of the wavefield as \texttt{u = TimeFunction(name='u', grid=grid)}, which is
denoted symbolically as \texttt{u(t, x, y)}.

\subsubsection{Spatial discretization}
The symbolic nature of the function objects allows the automatic derivation of
discretized finite difference expressions for derivatives. Devito
\texttt{Function} objects provide a set of shorthand notations that allow users
to express, for example, $\frac{d \vec{u}[t, x, y, z]}{d x}$ as \texttt{u.dx}
and $\frac{d^2 \vec{u}[t, x, y, z]}{d x^2}$ as \texttt{u.dx2}.  Moreover, the
discrete Laplacian, defined in three dimensions as $\Delta \vec{u}[t, x, y, z]
= \frac{d^2 \vec{u}[t, x, y, z]}{d x^2} + \frac{d^2 \vec{u}[t, x, y, z]}{d y^2}
+ \frac{d^2 \vec{u}[t, x, y, z]}{d z^2}$ can be expressed in shorthand simply
as \texttt{u.laplace}. The shorthand expression \texttt{u.laplace} is agnostic
to the number of spatial dimensions and may be used for two or
three-dimensional problems.

The discretization of the spatial derivatives can be defined for any order. In
the most general case, we can write the spatial discretization in the $x$
direction of order $k$ (and equivalently in the $y$ and $z$ direction) as:

\begin{equation}\label{eq:fdSpace}
\begin{aligned}
  \frac{\partial^2 \vec{u}[t, x, y, z]}{\partial x^2} &= \frac{1}{h_x^2} \sum_{j=0}^{\frac{k}{2}}
  \left[\alpha_j (\vec{u}[t, x + j h_x, y, z] + \vec{u}[t, x - j h_x, y, z]) \right],
\end{aligned}
\end{equation}
where $h_x$ is the discrete grid spacing for the dimension $x$, the
constants $\alpha_j$ are the coefficients of the finite
difference scheme and the spatial discretization error is of order
$O(h_x^k)$.

\subsubsection{Temporal discretization}

We consider here a second-order time discretization for the
acoustic wave equation, as higher order time discretization requires us to rewrite the PDE
\citep{Kimapnum}. The discrete second-order time derivative with this
scheme can be derived from the Taylor expansion of the discrete wavefield
$\vec{u}(t, x, y, z)$ as:
\begin{equation}\label{eq:fdTime}
  \frac{d^2 \vec{u}[t, x, y, z]}{d t^2} = \frac{\vec{u}[t+\Delta t, x, y, z]
    - 2 \vec{u}[t, x, y, z] + \vec{u}[t-\Delta t, x, y, z]}{\Delta t^2}.
\end{equation}
In this expression, $\Delta t$ is the size of a discrete time step. The
discretization error is $O(\Delta t^2)$ (second order in time) and will be
verified in Sec.~\ref{sec:tests}.

\begin{figure} \raggedright
\fbox{
  \includegraphics[width=.70\linewidth]{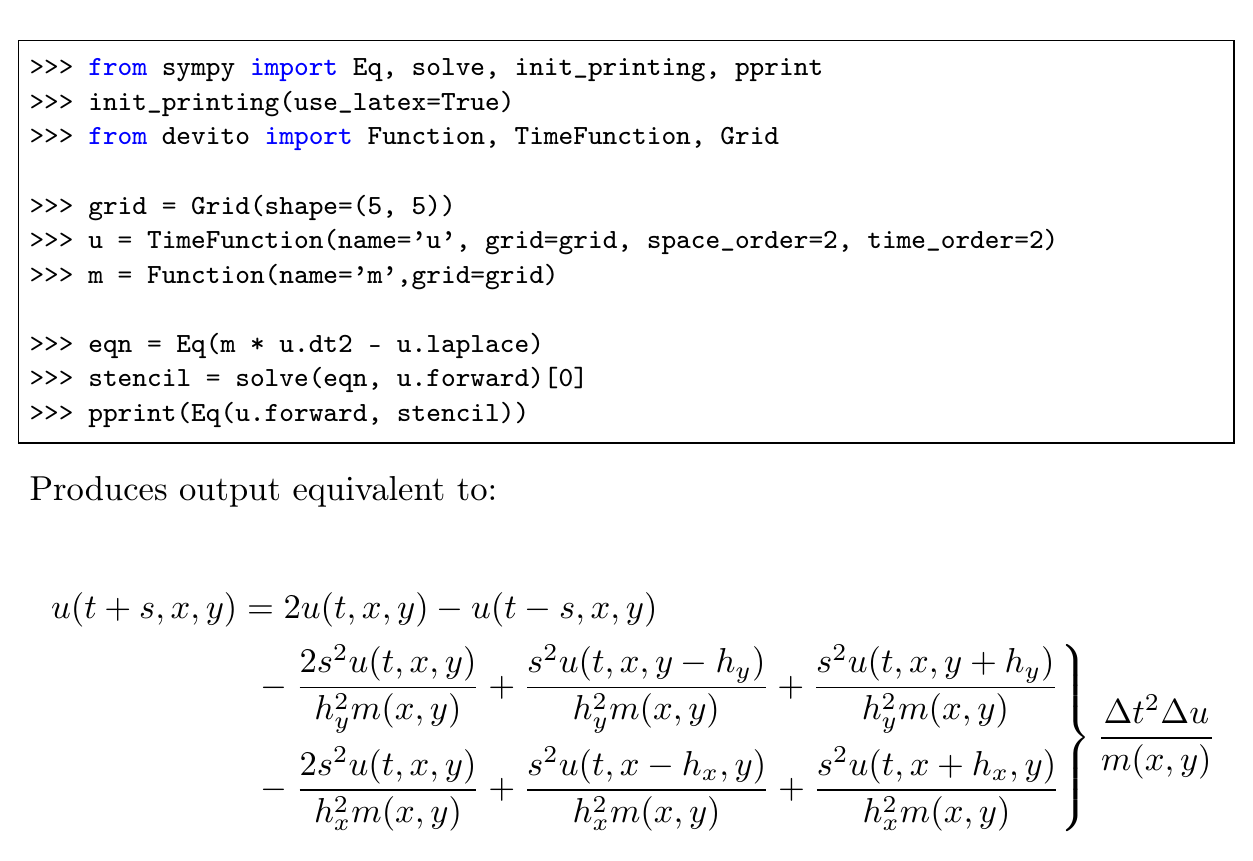}
  }
  \caption{Example code defining the two-dimensional wave equation
    without damping using Devito symbols and symbolic processing
    utilities from \Sympy. Assuming $h_{x} = \Delta x$, $h_{y} =
    \Delta y$ and $s = \Delta t$ the output is equivalent to
    Eq.~\ref{eq:acoustic} without the source term $\vec{q}_s$.}
  \label{lst:WE}
\end{figure}

Following the convention used for spatial derivatives, the above expression can
be automatically generated using the shorthand expression \texttt{u.dt2}.
Combining the temporal and spatial derivative notations, and ignoring the
source term $q$, we can now define the wave propagation component of
Eq.~\ref{eq:acoustic} as a symbolic expression via \texttt{Eq(m * u.dt2 -
u.laplace, 0)} where \texttt{Eq} is the \Sympy representation of an equation.
In the resulting expression, all spatial and temporal derivatives are expanded
using the corresponding finite difference terms. To define the propagation of
the wave in time, we can now rearrange the expression to derive a stencil
expression for the forward stencil point in time, $\vec{u}(t+\Delta t, x, y,
z)$, denoted by the shorthand expression \texttt{u.forward}. The forward
stencil corresponds to the explicit Euler time-stepping that updates the next
time-step \texttt{u.forward} from the two previous ones \texttt{u} and
\texttt{u.backward} (Eq.~\ref{eq:eulerfwd}). We use the \Sympy utility
\texttt{solve} to automatically derive the explicit time-stepping scheme, as
shown in Fig.~\ref{lst:WE} for the second order in space discretization.

\begin{equation}\label{eq:eulerfwd}
  \vec{u}[t+\Delta t, x, y, z] =  2\vec{u}[t, x, y, z] - \vec{u}[t-\Delta t, x, y, z] + \frac{\Delta t^2}{\vec{m}[x, y, z]} \Delta\vec{u}[t, x, y, z].
\end{equation}

The iteration over time to obtain the full solution is then generated by the
Devito compiler from the time dimension information.  Solving the wave-equation
with the above explicit Euler scheme is equivalent to a linear system
$\mathbf{A}(\vec{m})\vec{u}=\vec{q}_s$ where the vector $\vec{u}$ is the
discrete wavefield solution of the discrete wave-equation, $\vec{q}_s$ is the
source term and $\mathbf{A}(\vec{m})$ is the matrix representation of the
discrete wave-equation. From Eq.~\ref{eq:eulerfwd} we can see that the matrix
$\mathbf{A}(\vec{m})$ is a lower triangular matrix that reflects the
time-marching structure of the stencil. Simulation of the wavefield is
equivalent to a forward substitution (solve row by row from the top) on the
lower triangular matrix $\mathbf{A}(\vec{m})$. Since we do not consider complex
valued PDEs, the adjoint of $\mathbf{A}(\vec{m})$ is equivalent to its
transpose denoted as $\mathbf{A}^\top(\vec{m})$ and is an upper triangular
matrix. The solution $\vec{v}$ of the discrete adjoint wave-equation
$\mathbf{A}(\vec{m})^\top\vec{v}=\vec{q}_a$ for an adjoint source $\vec{q}_a$
is equivalent to a backward substitution (solve from the bottom row to top row) on
the upper triangular matrix $\mathbf{A}(\vec{m})^\top$ and is simulated
backward in time starting from the last time-step. These matrices are never
explicitly formed, but are instead matrix free operators with implicit
implementation of the matrix-vector product,
$\vec{u}=\mathbf{A}(\vec{m})^{-1}\vec{q}_s$ as a forward stencil. The stencil
for the adjoint wave-equation in this self-adjoint case would simply be
obtained with \texttt{solve(eqn, u.backward)} and the compiler will detect the
backward-in-time update.

\subsubsection{Boundary conditions}
\label{sec:boundary}

The field recorded data is measured on a wavefield that propagates in an
infinite domain. However, solving the wave equation in a discrete infinite
domain is not feasible with finite differences. In order to mimic an infinite
domain, Absorbing Boundary Conditions (ABC) or Perfectly Matched Layers (PML)
are necessary \citep{Clayton1529}. These two methods allow the approximation of
the wavefield as it is in an infinite medium by damping and absorbing the waves
within an extra layer at the limit of the domain to avoid unnatural reflections
from the edge of the discrete domain.

The least computationally expensive method is the Absorbing Boundary
Condition that adds a single damping mask in a finite layer around
the physical domain. This absorbing condition can be included in the
wave-equation as:

\begin{equation}\label{eq:fdWE-abc}
\begin{aligned}
  \vec{m}[x,y,z] \frac{d^2 \vec{u}[t, x, y, z]}{d t^2} -
   \Delta \vec{u}[t, x, y, z] +
  \vec{\eta}[x,y,z] \frac{d \vec{u}[t, x, y, z]}{d t} &=0.\\
\end{aligned}
\end{equation}

The $\vec{\eta}[x,y,z]$ parameter is equal to $0$ inside the physical domain
and increasing from inside to outside within the damping layer. The dampening
parameter $\vec{\eta}$ can follow a linear or exponential curve depending on the
frequency band and width of the dampening layer. For methods based on more
accurate modelling, for example in simulation-based acquisition design
\citep{doi:10.1190/geo2010-0231.1, wason2016GEOPctl, doi:10.1190/1.3008547,
kumar2015sss}, a full implementation of the PML will be necessary to avoid weak
reflections from the domain limits.

\subsubsection{Sparse point interpolation}
\label{sec:interpolation}

Seismic inversion relies on data fitting algorithms, hence we need to
support sparse operations such as source injection and wavefield ($\vec{u}[t,
x, y, z]$) measurement at arbitrary grid locations. Both
operations occur at sparse domain points, which do not necessarily align with
the logical cartesian grid used to compute the discrete solution $\vec{u}(t, x, y,
z)$. Since such operations are not captured by the finite differences
abstractions for implementing PDEs, Devito implements a secondary high-level
representation of sparse objects ~\citep{Lange2017scipy} to create a set of
\Sympy expressions that perform polynomial interpolation within the containing
grid cell from pre-defined coefficient matrices.

The necessary expressions to perform interpolation and injection are
automatically generated through a dedicated symbol type,
\texttt{SparseFunction}, which associates a set of coordinates with the symbol
representing a set of non-aligned points. For examples, the syntax
\texttt{p.interpolate(expr)} provided by a \texttt{SparseFunction p} will
generate a symbolic expressions that interpolates a generic expression
\texttt{expr} onto the sparse point locations defined by \texttt{p}, while
\texttt{p.inject(field, expr)} will evaluate and add \texttt{expr} to each
enclosing point in \texttt{field}. The generated \Sympy expressions are passed
to Devito \texttt{Operator} objects alongside the main stencil expression to be
incorporated into the generated C kernel code. A complete setup of the acoustic
wave equation with absorbing boundaries, injection of a source function and
measurement of wavefields via interpolation at receiver locations can be found
in Sec.~\ref{sec:acoustic_forward}.

\section{Seismic modeling and inversion}
\label{sec:seismic}

Seismic inversion methods aim to reconstruct physical parameters or an image of
the earth's subsurface from multi-experiment field measurements. For this
purpose, a wave is generated at the ocean surface that propagates through to the
subsurface and creates reflections at the discontinuities of the medium. The
reflected and transmitted waves are then captured by a set of hydrophones that
can be classified as either moving receivers (cables dragged behind a source
vessel) or static receivers (ocean bottom nodes or cables). From the acquired
data, physical properties of the subsurface such as wave speed or density can
be reconstructed by minimizing the misfit between the recorded measurements and
the numerically modelled seismic data.

\subsection{Full-Waveform Inversion}
\label{sec:fwi}

Recovering the wave speed of the subsurface from surface seismic measurements
is commonly cast into a non-linear optimization problem called full-waveform
inversion (FWI). The method aims at recovering an accurate model of the
discrete wave velocity, $\vec{c}$, or alternatively, the square slowness of the
wave, $\vec{m} = \frac{1}{\vec{c}^2}$(not an overload), from a given
set of measurements of the pressure wavefield $\vec{u}$. \citet{LionsJL1971, Tarantola,
Virieux, haber10TRemp} shows that this can be expressed as a
PDE-constrained optimization problem. After elimination of the PDE constraint,
the reduced objective function is defined as:

\begin{equation}
\begin{aligned}\label{eq:AS}
  \mathop{\hbox{minimize }}_{\vec{m}} \hspace*{.2cm} & \Phi_s(\vec{m}) =\frac{1}{2}\left\lVert\mathbf{P}_r
  \vec{u} - \vec{d}\right\rVert_2^2 \ \ \
  \mathop{\hbox{with: }} & \vec{u} = \mathbf{A}(\vec{m})^{-1} \mathbf{P}_s^T \vec{q}_s,
\end{aligned}
\end{equation}

where $\mathbf{P}_r$ is the sampling operator at the receiver locations,
$\mathbf{P}_s^T$ ($^T$ is the transpose or adjoint) is the injection operator
at the source locations, $\mathbf{A}(\vec{m})$ is the operator representing the
discretized wave equation matrix, $\vec{u}$ is the discrete synthetic pressure
wavefield, $\vec{q}_s$ is the corresponding pressure source and $\vec{d}$ is
the measured data. While we consider the acoustic isotropic wave equation for
simplicity here, in practice, multiple implementations of the wave equation
operator $\vec{A}(\vec{m})$ are possible depending on the choice of physics. In
the most advanced case, $\vec{m}$ would not only contain the square slowness
but also anisotropic or orthorhombic parameters.

To solve this optimization problem with a gradient-based method, we use the
adjoint-state method to evaluate the gradient \citep{PlessixASFWI,
haber10TRemp}:

\begin{align}\label{eq:Grad}
 \nabla\Phi_s(\vec{m})=\sum_{\vec{t} =1}^{n_t}\vec{u}[\vec{t}] \vec{v}_{tt}[\vec{t}] =\mathbf{J}^T\delta\vec{d}_s,
\end{align}

where $n_t$ is the number of computational time steps, $\delta\vec{d}_s =
\left(\mathbf{P}_r \vec{u} - \vec{d} \right)$ is the data residual (difference
between the measured data and the modeled data), $\mathbf{J}$ is the Jacobian
operator and $\vec{v}_{tt}$ is the second-order time derivative of the adjoint
wavefield that solves:

\begin{align}\label{eq:Adj}
 \mathbf{A}^T(\vec{m}) \vec{v} = \mathbf{P}_r^T \delta\vec{d}_s.
\end{align}

The discretized adjoint system in Eq.~\ref{eq:Adj} represents an upper
triangular matrix that is solvable by modelling wave propagation backwards in
time (starting from the last time step). The adjoint state method, therefore,
requires a wave equation solve for both the forward and adjoint wavefields to
compute the gradient. An accurate and consistent adjoint model for the solution
of the optimization problem is therefore of fundamental importance.

\subsection{Acoustic forward modelling operator} \label{sec:acoustic_forward}

\begin{figure}\raggedright
  \includegraphics[scale=1.0]{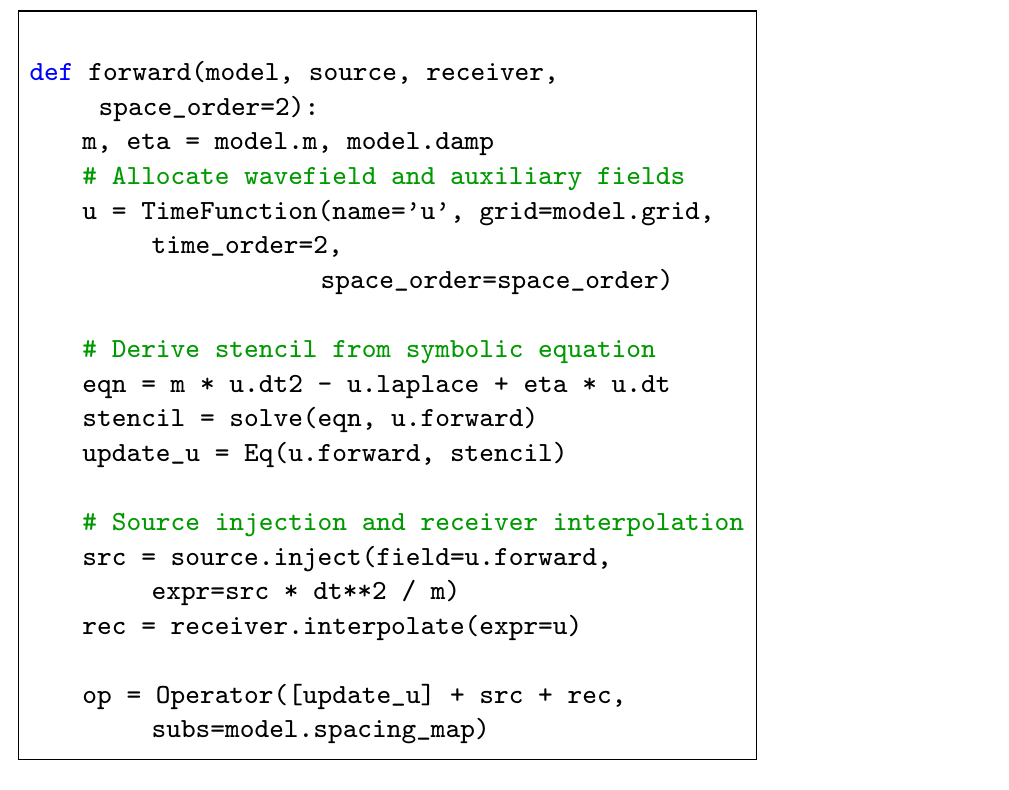}
  \caption{Example definition of a forward operator.}
  \label{lst:forward}
\end{figure}

We consider the acoustic isotropic wave-equation parameterized in terms of
slowness $\vec{m}[x,y,z]$ with zero initial conditions assuming
the wavefield does not have any energy before zero time. We define an
additional dampening term to mimic an infinite domain (see Sec.~\ref{sec:boundary}). At the limit of the domain,
the zero Dirichlet boundary condition is satisfied as the solution is considered to be
fully damped at the limit of the computational domain. The PDE is defined in
Eq.~\ref{eq:fdWE-abc}. Figure \ref{lst:forward} demonstrates the complete set
up of the acoustic wave equation with absorbing boundaries, injection of a
source function and sampling wavefields at receiver locations. The shape of the
computational domain is hereby provided by a utility object \texttt{model},
while the damping term $\eta\frac{d \vec{u}[x,y,z,t]}{dt}$ is implemented via a
utility symbol \texttt{eta} defined as a \texttt{Function} object. It is
important to note that the discretization order of the spatial derivatives is
passed as an external parameter \texttt{order} and carried as meta-data by the
wavefield symbol \texttt{u} during construction, allowing the user to freely
change the underlying stencil order.

The main (PDE) stencil expression to update the state of the wavefield is
derived from the high-level wave equation expression \texttt{eqn = u.dt2 -
u.laplace + damp*u.dt} using \Sympy utilities as demonstrated before in
Fig.~\ref{lst:WE}. Additional expressions for the injection of the wave source
via the \texttt{SparseFunction} object \texttt{src} are then generated for the
forward wavefield, where the source time signature is discretized onto the
computational grid via the symbolic expression \texttt{src * dt**2 / m}. The
weight $\frac{dt^2}{m}$ is derived from rearranging the discretized wave
equation with a source as a right-hand-side similarly to the Laplacian in
Eq.~\ref{eq:eulerfwd}. A similar expression to interpolate the current state of
the wavefield at the receiver locations (measurement points) is generated
through the \texttt{receiver} symbol. The combined list of stencils, a sum in
Python that adds the different expressions that update the wavefield at
the next time step, inject the source and interpolate at the receivers, is then
passed to the \texttt{Operator} constructor alongside a definition of the
spatial and temporal spacing $h_x, h_y, h_z, \Delta t$ provided by the
\texttt{model} utility. Devito then transforms this list of stencil expressions
into loops (inferred from the symbolic Functions), replaces all necessary
constants by their values if requested, prints the generated C code and
compiles it. The operator is finally callable in Python with
\texttt{op.apply()}.

A more detailed explanation of the seismic setup and parameters such as the source and receiver terms in Fig.~\ref{lst:forward} is covered in \citet{TLE1}.

\subsection{Discrete adjoint wave-equation and FWI gradient}

\begin{figure}
  \begin{minipage}[t][][t]{.49\textwidth}
    \includegraphics[scale=1.0]{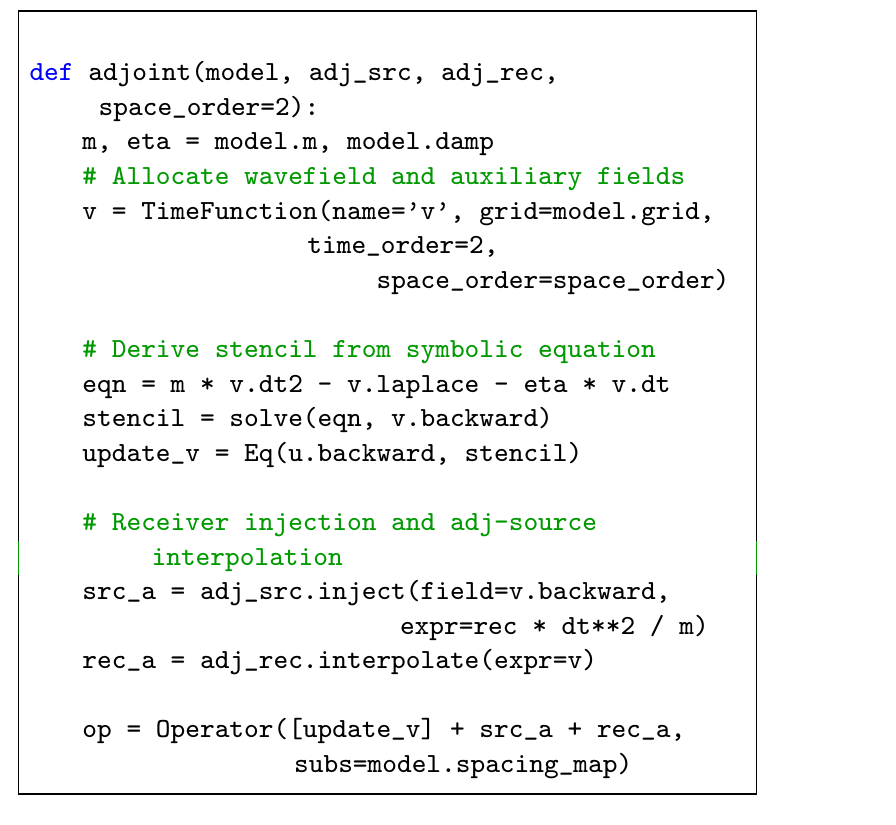}
    \caption{Example definition of an adjoint operator.}
    \label{lst:adjoint}
  \end{minipage}
  \begin{minipage}[t][][t]{.49\textwidth}
    \includegraphics[scale=1.0]{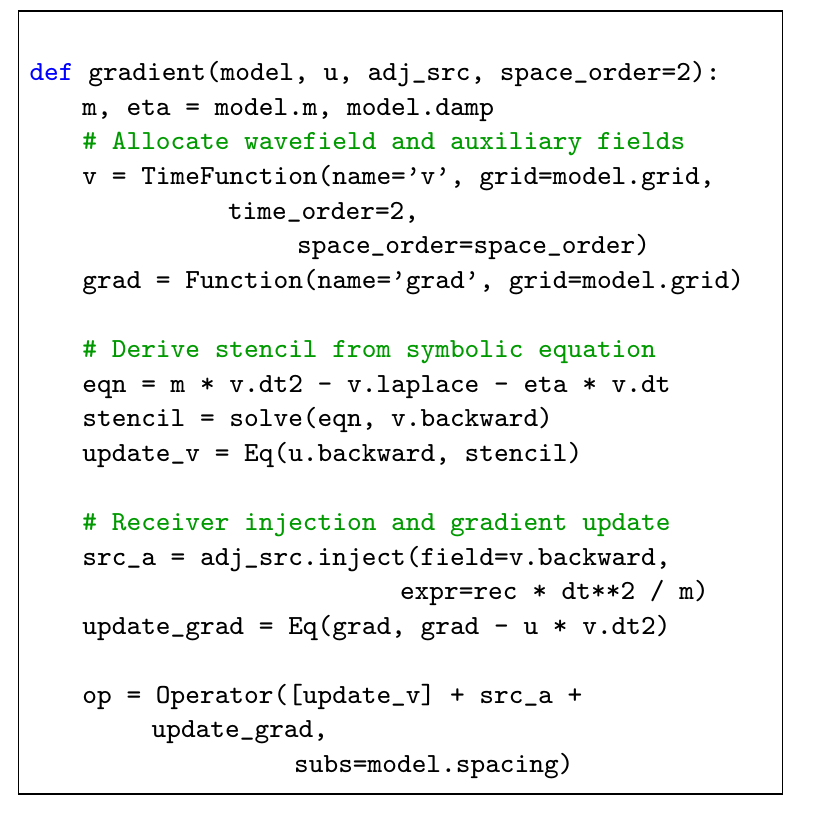}
    \caption{Example definition of a gradient operator.}
    \label{lst:gradient}
  \end{minipage}
\end{figure}

To create the adjoint that pairs with the above forward modeling propagator we
can make use of the fact that the isotropic acoustic wave equation is
self-adjoint. This entails that for the implementation of the forward wave
equation \texttt{eqn}, shown in Fig.~\ref{lst:adjoint}, only the sign of the
damping term needs to be inverted, as the dampening time-derivative has to be
defined in the direction of propagation ($\frac{\partial}{\partial n(t)}$). For
the PDE stencil, however, we now rearrange the stencil expression to update the
backward wavefield from the two next time steps as $\vec{v}[t - \Delta t,x,y,z]
= f(\vec{v}[t,x,y,z], \vec{v}[t + \Delta t,x,y,z])$. Moreover, the role of the
sparse point symbols has changed (Eq.~\ref{eq:Adj}), so that we now inject
time-dependent data at the receiver locations (\texttt{adj\_src}), while
sampling the wavefield at the original source location (\texttt{adj\_rec}).

Based on the definition of the adjoint operator, we can now define a similar
operator to update the gradient according to Eq.~\ref{eq:Grad}. As shown in
Fig.~\ref{lst:gradient}, we can replace the expression to sample the wavefield
at the original source location with an accumulative update of the gradient
field \texttt{grad} via the symbolic expression \texttt{Eq(grad, grad - u *
v.dt2)}.

To compute the gradient, the forward wavefield at each time step must be
available which leads to significant memory requirements. Many methods exist to
tackle this memory issue, but all come with their advantages and disadvantages.
For instance, we implemented optimal
checkpointing with the library Revolve \citep{Griewank:2000:ARI:347837.347846} in
Devito to drastically reduce the memory cost by only saving a partial time
history and recomputing the forward wavefield when needed~\citep{kukreja2018high}. The memory reduction
comes at an extra computational cost as optimal checkpointing requires
$~log(n_t)+2$ extra PDE solves. Another method is boundary wavefield
reconstruction~\citep{McMechan, Mittet, RaknesR45} that saves the wavefield
only at the boundary of the model, but still requires us to recompute the
forward wavefield during the back-propagation. This boundary method has a
reduced memory cost but necessitates the computation of the forward wavefield
twice (one extra PDE solve), once to get the data than a second time from the
boundary values to compute the gradient.

\subsection{FWI using Devito operators}

\begin{figure} \centering
  \begin{minipage}[t][][t]{.49\textwidth}
    \includegraphics[scale=1.0]{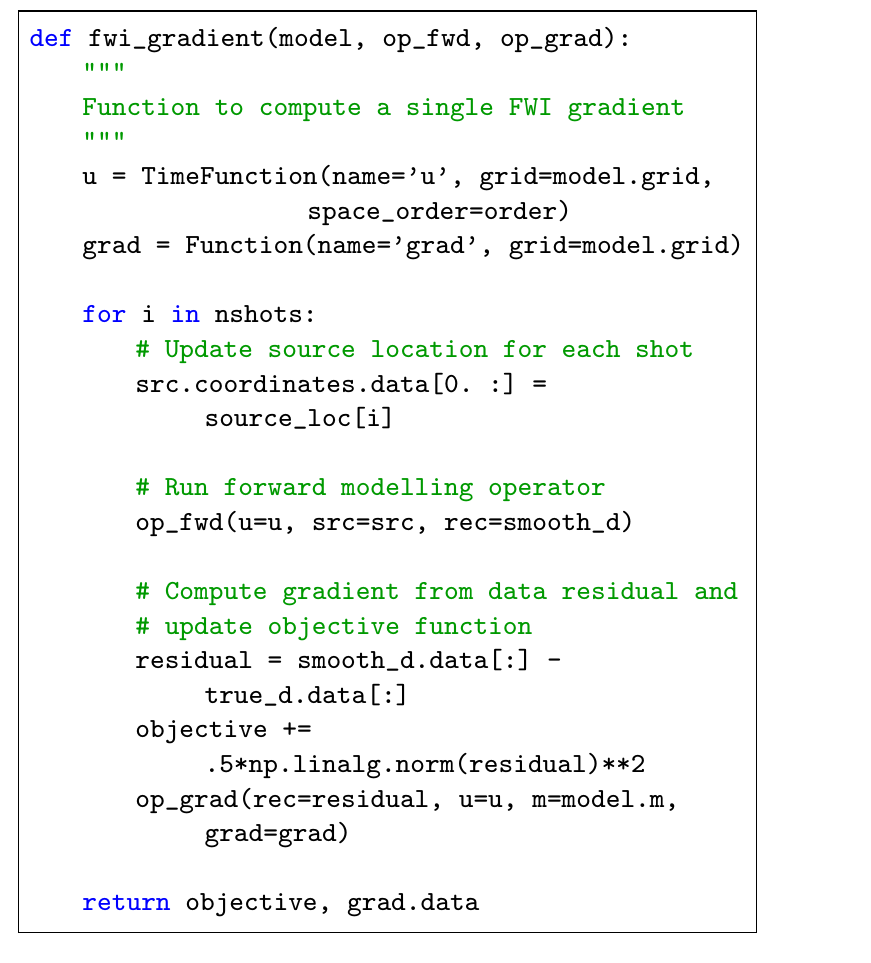}
    \caption{Definition of FWI gradient update.}
    \label{lst:fwi_gradient}
  \end{minipage}
  \begin{minipage}[t][][t]{.49\textwidth}
    \includegraphics[scale=1.0]{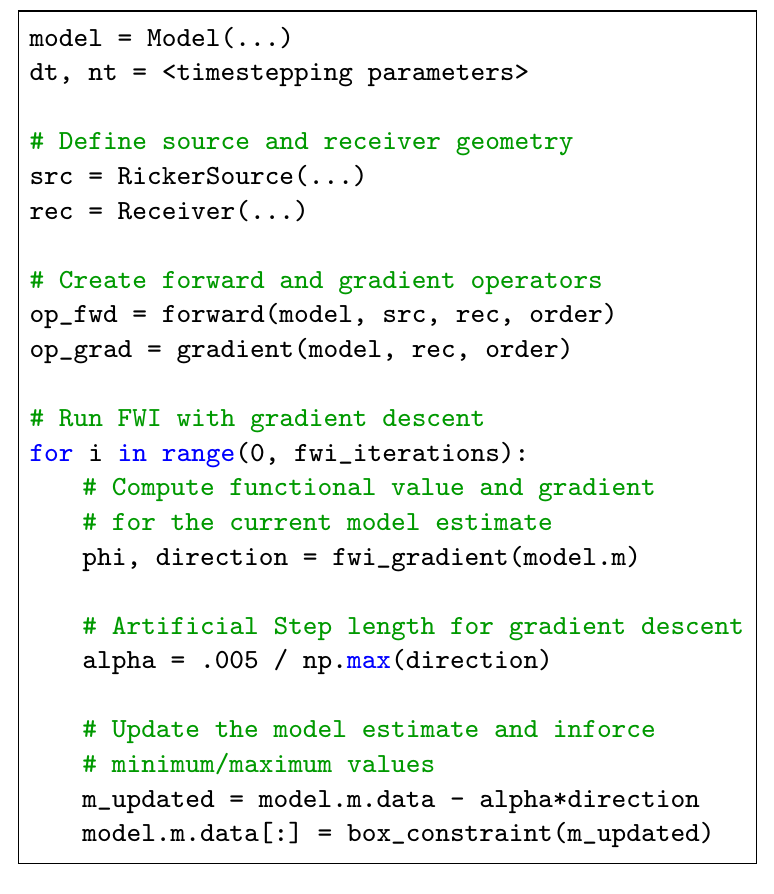}
    \caption{FWI algorithm with linesearch.}
    \label{lst:fwi}
  \end{minipage}
\end{figure}

At this point, we have a forward propagator to model synthetic data in
Fig.~\ref{lst:forward}, the adjoint propagator for Eq.~\ref{eq:Adj} and the FWI
gradient of Eq.~\ref{eq:Grad} in Fig.~\ref{lst:gradient}. With these three
operators, we show the implementation of the FWI objective and gradient with
Devito in  Fig.~\ref{lst:fwi}. With the forward and adjoint/gradient operator
defined for a given source, we only need to add a loop over all the source
experiments and the reduction operation on the gradients (sum the gradient for
each source experiment together). In practice, this loop over sources is where
the main task-based or MPI based parallelization happens. The wave-equation
propagator does use some parallelization with multithreading or domain
decomposition but that parallelism requires communication. The parallelism over source
experiment is task-based and does not require any communication between the
separate tasks as the gradient for each source can be computed independently
and reduced to obtain the full gradient. With the complete gradient summed over
the source experiments, we update the model with a simple fixed step length
gradient update \citep{Cauchy1847Methode}.

This FWI function in Fig.~\ref{lst:fwi_gradient} can then be included in any
black-box optimization toolbox such as \SciPy \texttt{optimize} to solve the
inversion problem Eq.\ref{eq:AS}. While black-box optimization methods
aim to minimize the objective, there are no guarantees they find a global minimum because the
objective is highly non-linear in $m$ and other more sophisticated methods are
required \citep{AWI, vanleeuwen2015IPpmp, peters2016cvp, witte2018fwip3}.

\section{Verification}\label{sec:tests}

Given the operators defined in Sec.~\ref{sec:devito} we now verify the
correctness of the code generated by the Devito compiler. We first verify that
the discretized wave equation satisfies the convergence properties defined by
the order of discretization, and secondly we verify the correctness of the
discrete adjoint and computed gradient.

\subsection{Numerical accuracy}
\label{sec:convergence}

The numerical accuracy of the forward modeling operator
(Fig.~\ref{lst:forward}) and the runtime achieved for a given spatial
discretization order and grid size are compared to the analytical solution of the wave
equation in a constant media. We define two measures of the accuracy that
compare the numerical wavefield in a constant velocity media to the analytical
solution:

\begin{itemize}
    \item \textbf{Accuracy versus size}, where we compare the obtained
        numerical accuracy as a function of the spatial sampling size (grid
        spacing).
    \item \textbf{Accuracy versus time}, where we compare the obtained
        numerical accuracy as a function of runtime for a given physical model
        (fixed shape in physical units, variable grid spacing).
\end{itemize}

The measure of accuracy of a numerical solution relies on a hypothesis that we
satisfy for these two tests:

\begin{itemize}
    \item The domain is large enough and the propagation time small enough to
        ignore boundary related effects, i.e. the wavefield never reaches the
        limits of the domain.
    \item The source is located on the grid and is a discrete approximation of
        the Dirac to avoid spatial interpolation errors. This hypothesis
        guarantees the existence of the analytical and numerical solution for
        any spatial discretization \citep{id2138}.
\end{itemize}

\paragraph{Convergence in time}

\begin{figure}
  \begin{minipage}[t][][t]{.49\textwidth}
    \includegraphics[width=\linewidth,trim={1.5cm 0 1.5cm 0}, clip]{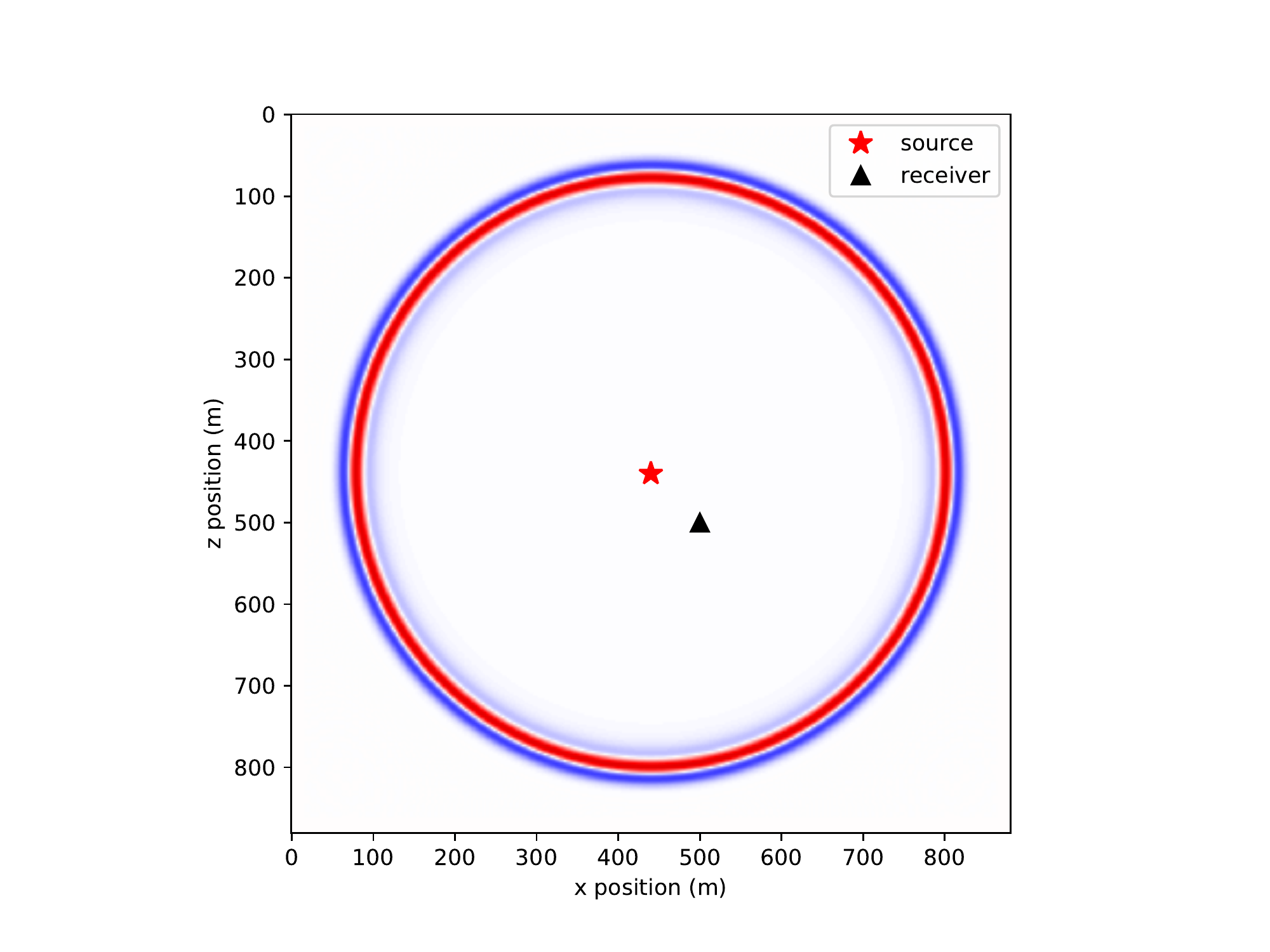}
  \end{minipage}
  \begin{minipage}[t][][t]{.49\textwidth}
    \includegraphics[width=\linewidth, trim={1.5cm 0 1.5cm 0}, clip]{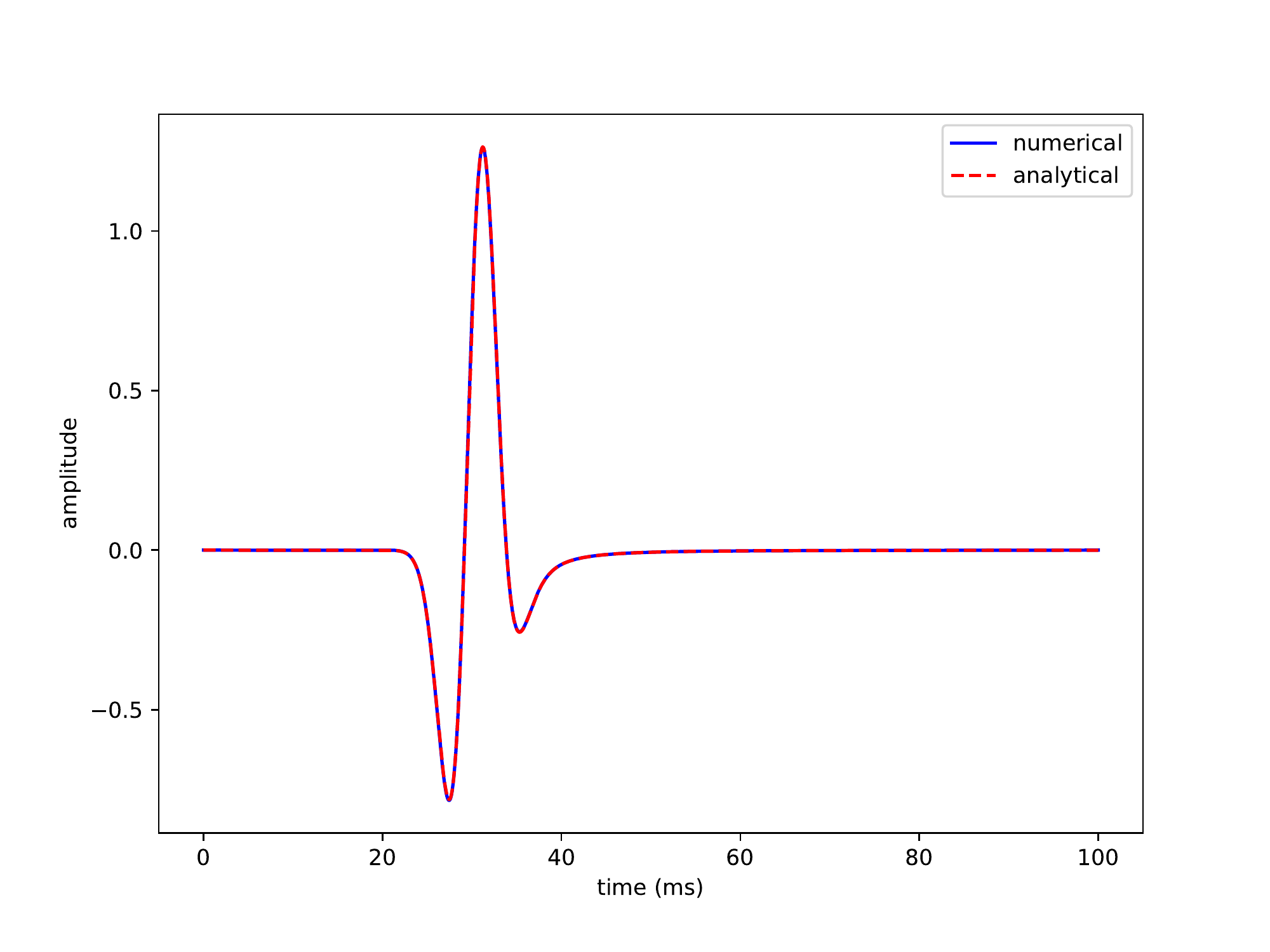}
  \end{minipage}
  \caption{Numerical wavefield for a constant velocity $dt=.1\unit{ms}$, $h=1\unit{m}$ and comparison with the analytical solution.}
  \label{fig:refanaly}
\end{figure}

We analyze the numerical solution against the analytical solution and verify that the
error between these two decreases at a second order rate as a function
of the time step size $\Delta t$. The velocity model is a $400\unit{m} \times 400\unit{m}$ domain with
a source at the center. We compare the numerical solution to the analytical
solution on Fig.~\ref{fig:refanaly}.

\begin{figure}
\centering
\includegraphics[width=0.9\linewidth]{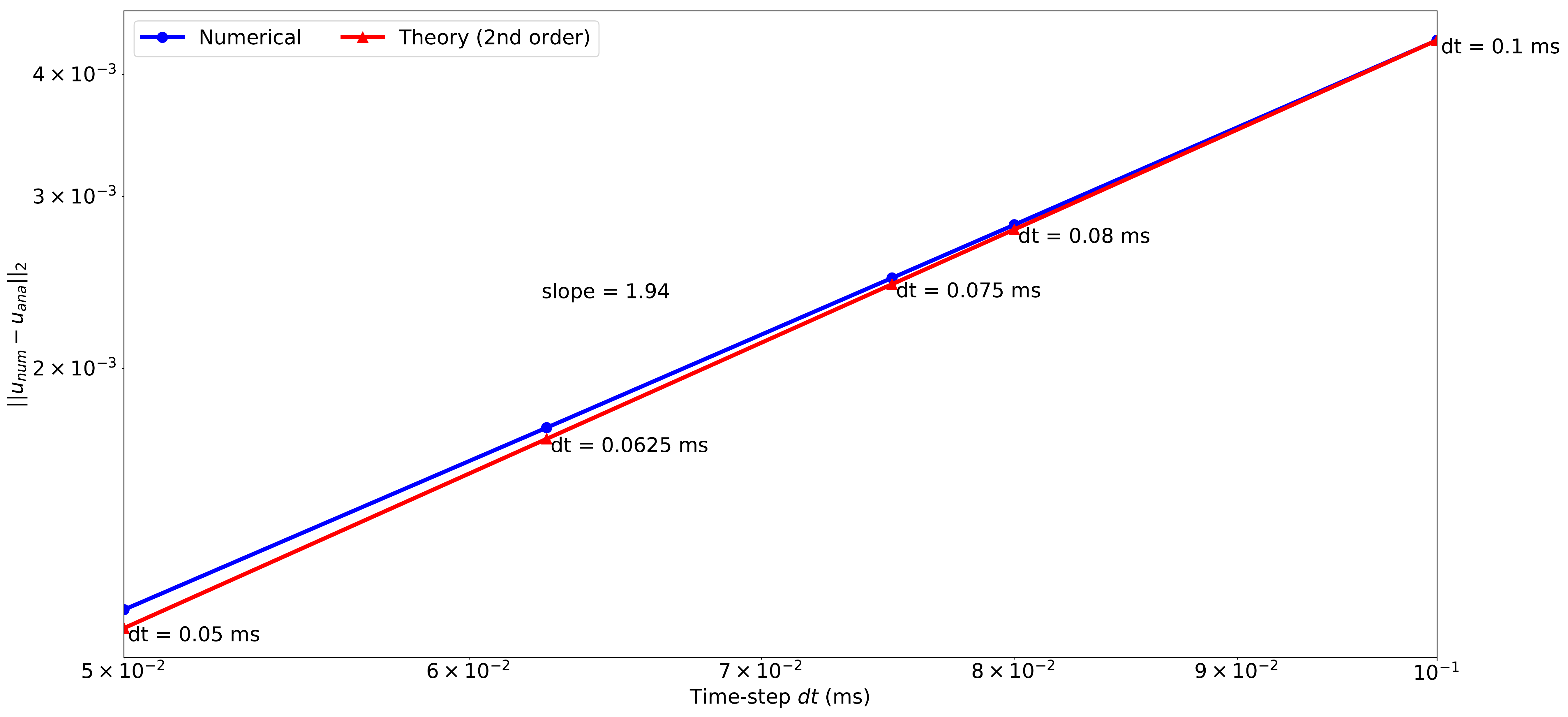}
\caption{Time discretization convergence analysis for a fixed grid, fixed
    propagation time (150ms) and varying time step values. The result is
    plotted in a logarithmic scale and the numerical convergence rate (1.94
    slope) shows that the numerical solution is accurate.}
\label{fig:ctime}
\end{figure}

The analytical solution is defined as~\citep{Watanabe2015}:
\begin{align}
u_s(r, t) &= \frac{1}{2\pi} \int_{-\infty}^{\infty} \{ -i \pi H_0^{(2)}\left(k r \right) q(\omega) e^{i\omega t} d\omega\} \\
r &= \sqrt{(x - x_{src})^2+(y - y_{src})^2},
\label{eq:analytical}
\end{align}
where $H_0^{(2)}$ is the Hankel function of second kind and  $q(\omega)$ is the
spectrum of the source function. As we can see on Fig.~\ref{fig:ctime} the
error decreases near quadratically with the size of the time step with a time convergence rate of slope of 1.94 in logarithmic scale that
matches the theoretical expectation from a second order temporal discretization.

\paragraph{Spatial discretization analysis}

\begin{figure}
\centering
\includegraphics[width=0.8\linewidth]{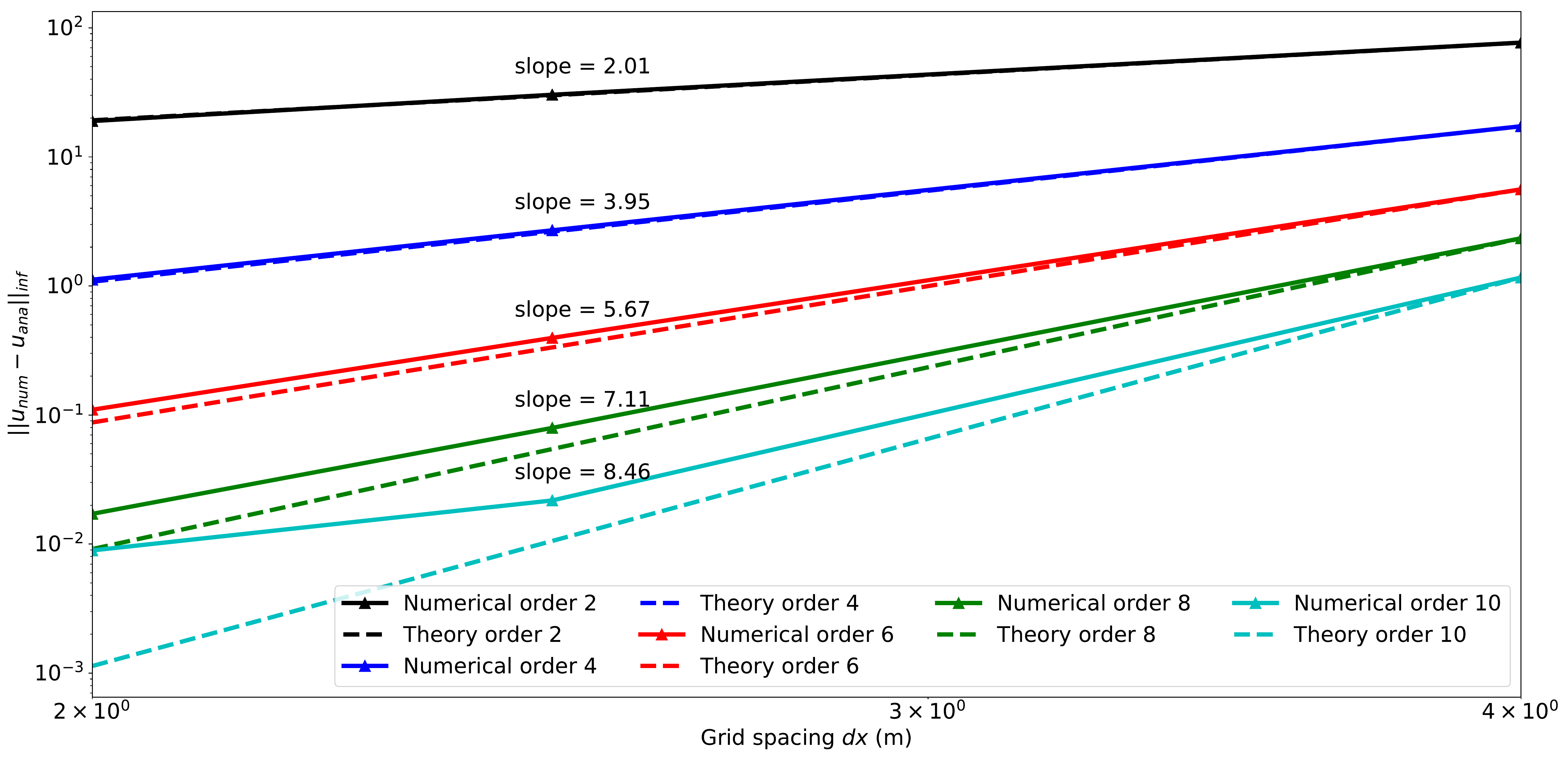}
\caption{Comparison of the numerical convergence rate of the spatial
    finite difference scheme with the theoretical convergence rate from the
    Taylor theory. The theoretical rates are the dotted line with the
    corresponding colors. The result is plotted in a logarithmic scale to
    highlight the convergence orders as linear slopes and the numerical
    convergence rates show that numerical solution is accurate.}
\label{fig:AvO}
\end{figure}

The spatial discretization analysis follows the same method as the temporal discretixzation analysis.
We model a wavefield for a fixed temporal setup with a small enough time-step to ensure negligeable
time discretization error ($dt=.00625ms$). We vary the grid spacing ($dx$) and spatial discretization order and the
and compute the error between the numerical and analytical solution.
The convergence rates should follow the theoretical rates defined in Eq.~\ref{eq:fdSpace}. In details, for a
$k^{th}$ order discretization in space, the error between the numerical and analytical solution should decrease as $O(dx^k)$.
The best way to look at the convergence results is to plot the error in logarithmic scale and verify that the error decrease linearly with slope $k$.
We show the convergence results on Fig.~\ref{fig:AvO}. The numerical convergence rates follow the theoretical ones for every
tested order $k=2, 4, 6, 8$ with the exception of the $10^{th}$ order for small
grid size. This is mainly due to reaching the limits of the numerical accuracy
and a value of the error on par with the temporal discretization error. This
behavior for high order and small grids is however in accordance with the
literature as in in~\citet{Wang1090158}.

The numerical slopes obtained and displayed on Fig.~\ref{fig:AvO} demonstrate
that the spatial finite difference follows the theoretical errors and converges
to the analytical solution at the expected rate. These two convergence results
(time and space) verify the accuracy and correctness of the symbolic
discretization with Devito. With this validated simulated wavefield, we can now
verify the implementation of the operators for inversion.

\subsection{Propagators verification for inversion}

We concentrate now on two tests, namely the adjoint test (or dot test) and the
gradient test. The adjoint state gradient of the objective function defined in
Eq.~\ref{eq:Grad} relies on the solutions of the forward and adjoint wave
equations, therefore, the first mandatory property to verify is the exact
derivation of the discrete adjoint wave equation. The mathematical test we use
is the standard adjoint property or dot-test:

\begin{align}
 \text{for any random } \vec{x} \in \text{span}(\vec{P}_s\vec{A}(\vec{m})^{-T}\vec{P}_r^{-T}), \vec{y} \in \text{span}(\vec{P}_r\vec{A}(\vec{m})^{-1}\vec{P}_s^{-T})\nonumber \\
 \frac{<\vec{P}_r\vec{A}(\vec{m})^{-1}\vec{P}_s^{-T}\vec{x},\vec{y}> - <\vec{x},\vec{P}_s\vec{A}(\vec{m})^{-T}\vec{P}_r^{-T}\vec{y}>}{<\vec{P}_r\vec{A}(\vec{m})^{-1}\vec{P}_s^{-T}\vec{x},\vec{y}>} = 0.0.
\label{eq:AdjTestPrRatio}
\end{align}

The adjoint test is also individually performed on the source/receiver
injection/interpolation operators in the Devito tests suite. The results,
summarized in Tables~\ref{tab:adjtest2D} and~\ref{tab:adjtest3D} with $\textbf{F}
=\vec{P}_r\vec{A}(\vec{m})^{-1}\vec{P}_s^{-T}$, verify the correct
implementation of the adjoint operator for any order in both 2D and 3D. We
observe that the discrete adjoint is accurate up to numerical precision for any
order in 2D and 3D with an error of order $1e-16$. In combination with the
previous numerical analysis of the forward modeling propagator that guarantees
that we solve the wave equation, this result verifies that the adjoint
propagator is the exact numerical adjoint of the forward propagator and that it
implements the adjoint wave equation.

\begin{table}
\centering
\begin{minipage}[t]{.43\textwidth}
\begin{tabular}{cccccc}
Order & $<\textbf{F}\vec{x},\vec{y}>$ & $<\vec{x},\textbf{F}^{T}\vec{y}>$ &
           relative error \\
\hline
2nd order  &  7.9858e+05  &  7.9858e+05 & \textcolor{red}{0.0000e+00} \\
4th order  &  7.3044e+05  &  7.3044e+05 & \textcolor{red}{0.0000e+00} \\
6th order  &  7.2190e+05  &  7.2190e+05 & \textcolor{red}{4.8379e-16} \\
8th order  &  7.1960e+05  &  7.1960e+05 & \textcolor{red}{4.8534e-16} \\
10th order  &  7.1860e+05  &  7.1860e+05 & \textcolor{red}{3.2401e-16} \\
12th order  &  7.1804e+05  &  7.1804e+05 & \textcolor{red}{6.4852e-16} \\
\end{tabular}
\caption{Adjoint test for different discretization orders in 2D,
computed on a two layer model in double precision.}
\label{tab:adjtest2D}
\end{minipage}\hspace{1cm}
\begin{minipage}[t]{.43\textwidth}
\begin{tabular}{cccccc}
Order & $<\textbf{F}\vec{x},\vec{y}>$ & $<\vec{x},\textbf{F}^{T}\vec{y}>$ &
           relative error \\
\hline
2nd order  & 5.3840e+04  &  5.3840e+04 & \textcolor{red}{1.3514e-16} \\
4th order  & 4.4725e+04  &  4.4725e+04 & \textcolor{red}{3.2536e-16} \\
6th order  & 4.3097e+04  &  4.3097e+04 & \textcolor{red}{3.3766e-16} \\
8th order  & 4.2529e+04  &  4.2529e+04 & \textcolor{red}{3.4216e-16} \\
10th order  & 4.2254e+04  &  4.2254e+04 & \textcolor{red}{0.0000e+00} \\
12th order  & 4.2094e+04  &  4.2094e+04 & \textcolor{red}{1.7285e-16} \\
\end{tabular}
\caption{Adjoint test for different discretization orders in 3D,
computed on a two layer model in double precision.}
\label{tab:adjtest3D}
\end{minipage}
\end{table}

With the forward and adjoint propagators tested, we finally verify that the
Devito operator that implements the gradient of the FWI objective function
(Eq.~\ref{eq:Grad}, Fig.\ref{lst:gradient}) is accurate with respect to the
Taylor expansion of the FWI objective function. For a given velocity model and
associated squared slowness $\vec{m}$, the Taylor expansion of the FWI
objective function from Eq.~\ref{eq:AS} for a model perturbation $\vec{dm}$ and
a perturbation scale $h$ is:

\begin{align}
 &\Phi_s(\vec{m} + h \vec{dm}) = \Phi_s(\vec{m}) + \mathcal{O} (h) \nonumber\\
 &\Phi_s(\vec{m} + h \vec{dm}) = \Phi_s(\vec{m}) + h \langle\nabla\Phi_s(\vec{m}),\vec{dm}\rangle + \mathcal{O} (h^2).
\label{GrFWI}
 \end{align}
These two equations constitute the gradient test where we define a small model perturbation
$\vec{dm}$ and vary the value of $h$ between $10^{-6}$ and $10^{0}$ and compute
the error terms:
\begin{align}
 \epsilon_0 = &\Phi_s(\vec{m} + h \vec{dm}) - \Phi_s(\vec{m})\nonumber\\
 \epsilon_1 = &\Phi_s(\vec{m} + h \vec{dm}) - \Phi_s(\vec{m}) - h \langle\nabla\Phi_s(\vec{m}),\vec{dm}\rangle.
\label{GrFWI:test}
 \end{align}
We plot the evolution of the error terms as a function of the perturbation
scale $h$ knowing $\epsilon_0$ should be first order (linear with slope 1 in a
logarithmic scale) and $\epsilon_1$ should be second order (linear with slope 2
in a logarithmic scale). We executed the gradient test defined in
Eq.~\ref{GrFWI} in double precision with a $8^{th}$ order spatial
discretization. The test can be run for higher orders in the same manner but
since it has already been demonstrated that the adjoint is accurate for all
orders, the same results would be obtained.
\begin{figure}
\centering
\includegraphics[width=0.8\linewidth]{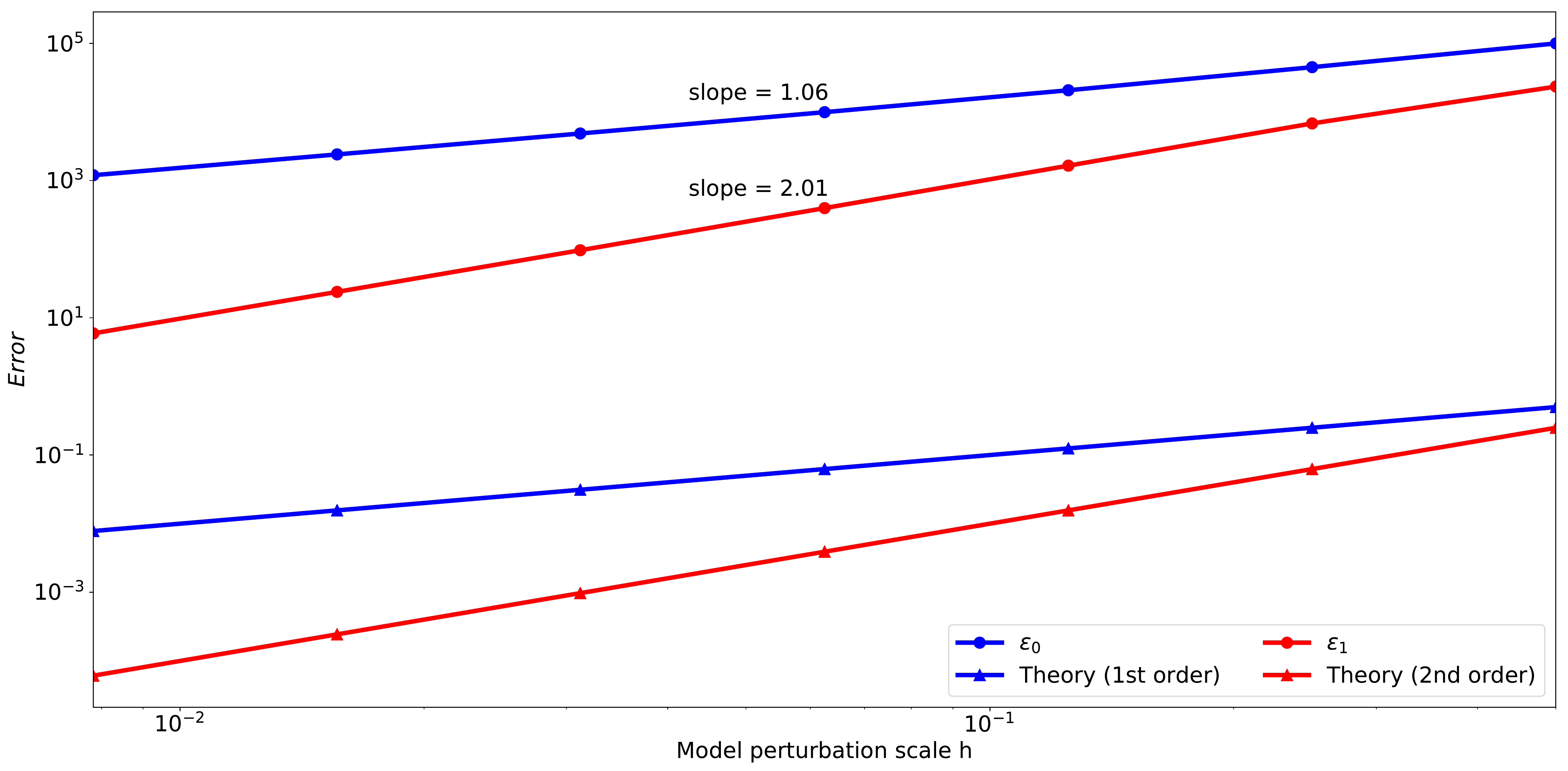}
\caption{Gradient test for the acoustic propagator. The first order (blue) and
    second order (red) errors are displayed in logarithmic scales to highlight
    the slopes. The numerical convergence order (1.06 and 2.01) show that we
    have a correct implementation of the FWI operators.}
\label{fig:Grad}
\end{figure}

In Fig.~\ref{fig:Grad}, the matching slope of the error term with the
theoretical $h$ and $h^2$ slopes from the Taylor expansion verifies the
accuracy of the inversion operators. With all the individual parts necessary
for seismic inversion, we now validate our implementation on a simple but
realistic example.

\subsection{Validation: Full-Waveform Inversion}

\begin{figure}  \centering
  \includegraphics[width=0.49\textwidth]{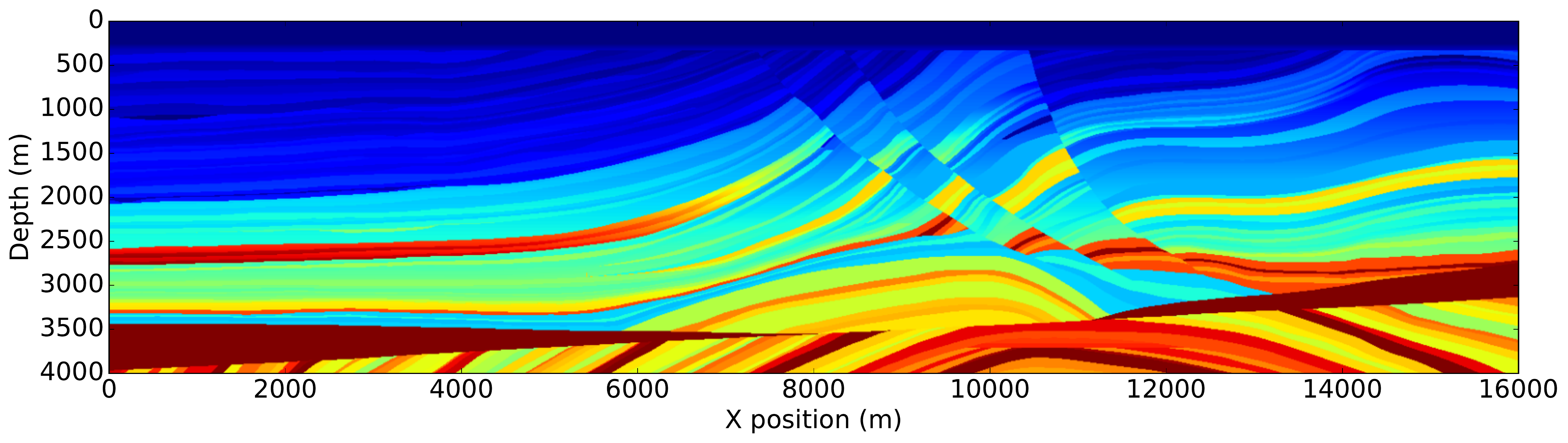}
  \includegraphics[width=0.49\textwidth]{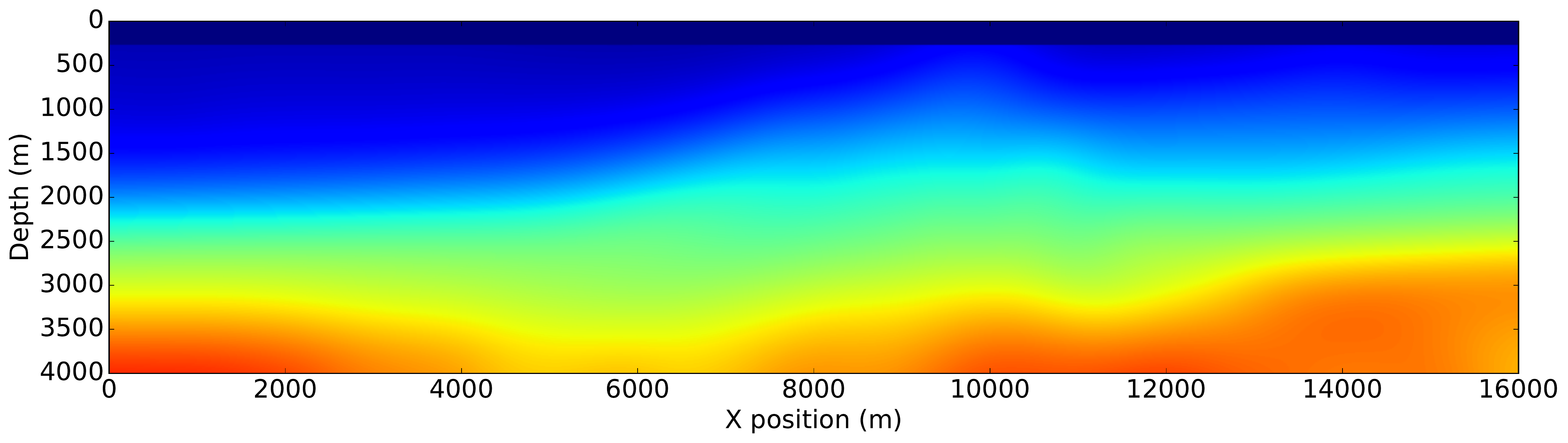} \\
  \includegraphics[width=0.49\textwidth]{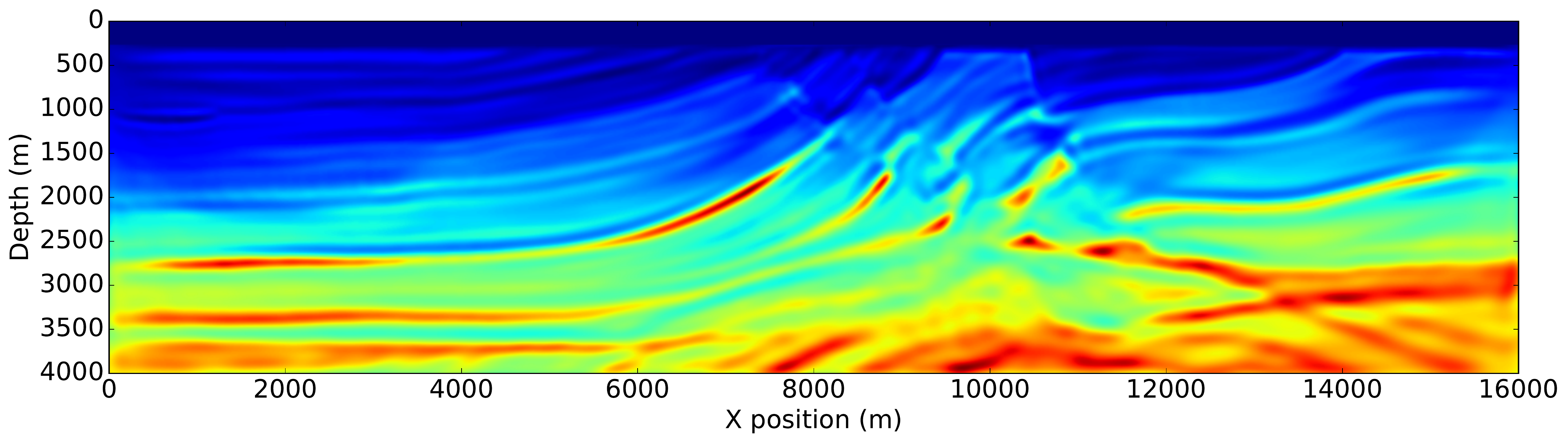}
  \includegraphics[width=0.49\textwidth]{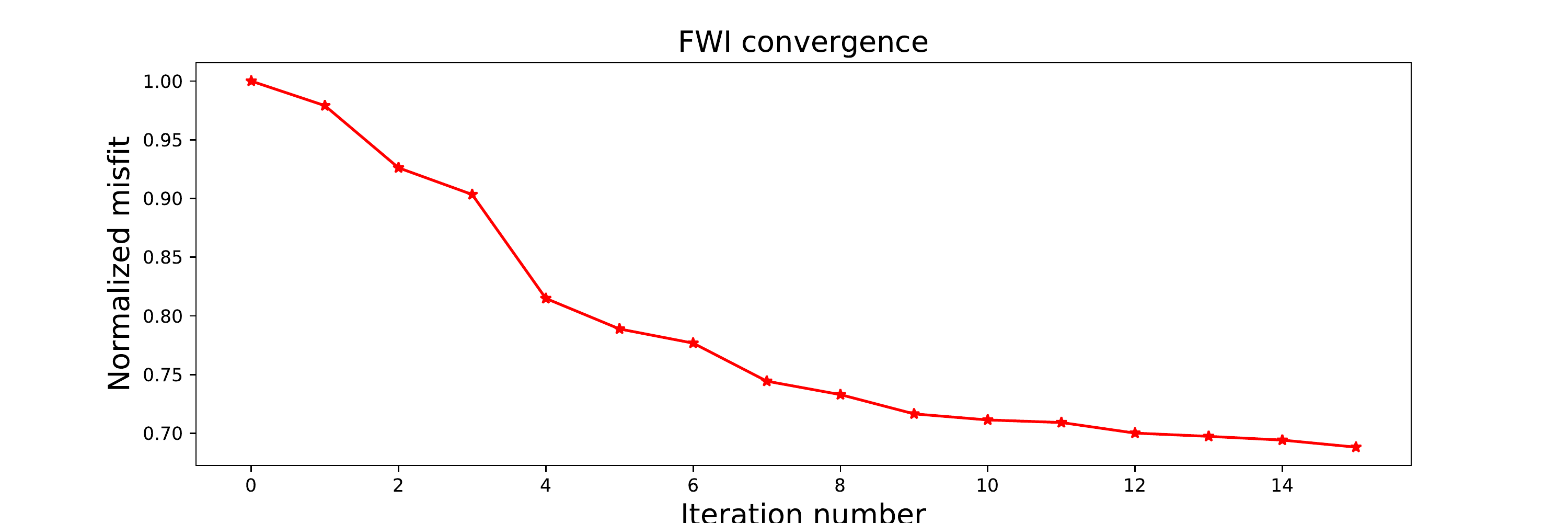}
  \caption{FWI on the acoustic Marmousi-ii model. The top-left plot is the
    true velocity model, the top-right is the initial velocity model,
    the bottom-left plot is the inverted velocity at the last iteration
    of the iterative inversion and the bottom-right plot is the convergence.}
  \label{fig:FWI}
\end{figure}

We show a simple example of FWI
Eq.~\ref{eq:Grad} on the Marmousi-ii model \citep{Versteeg927}. This result
obtained with the Julia interface to Devito JUDI \citep{witte2018fwip3,
Witte2018b} that provides high-level abstraction for optimization and linear
algebra. The model size is $4\unit{km}\times16\unit{km}$ discretized with a
$10\unit{m}$ grid in both directions. We use a $10\unit{Hz}$ Ricker wavelet
with $4\unit{s}$ recording. The receivers are placed at the ocean bottom
($210\unit{m}$ depth) every $10\unit{m}$. We invert for the velocity with all
the sources, spaced by $50\unit{m}$ at $10\unit{m}$ depth for a total of 300
sources. The inversion algorithm used is
minConf\_PQN\citep{SchmidtBergFriedlanderMurphy:2009}, an l-BFGS algorithm with
bounds constraints (minimum and maximum velocity values constraints).
While conventional optimization would run the algorithm to convergence,
this strategy is computationally not feasible for FWI. {\color{black} As each iteration requires two PDE solves per source $q_s$ (see adjoint state in Sec.~\ref{sec:seismic}),
 we can only afford a $\mathcal{O}(10)$ iterations in practice ($\mathcal{O}(10^4)$ PDE solves in total)}. In this example, we fix the number
of function evaluations to 20, which, with the line search, corresponds to 15 iteration.
The result is shown in Fig.~\ref{fig:FWI} and we can see that we obtain a good
reconstruction of the true model. More advanced algorithms and constraints will
be necessary for more complex problem such as less accurate initial model,
noisy data or field recorded data \citep{witte2018fwip3, peters2016cvp};
however the wave propagator would not be impacted, making this example a good
proof of concept for Devito.

This result highlights two main contributions of Devito. First, we provide PDE
simulation tools that allow easy and efficient implementation of inversion
operator for seismic problem and potentially any PDE constrained optimization
problem. As described in Sec.~\ref{sec:devito} and~\ref{sec:seismic}, we can
implement all the required propagators and the FWI gradient in a few lines in a
concise and mathematical manner. Second, as we obtained this results with
JUDI~\citep{witte2018alf}, a seismic inversion framework that provides a
high-level linear abstraction layer on top of Devito for seismic inversion,
this example illustrates that Devito is fully compatible with external
languages and optimizations toolboxes and allows users to use our symbolic DSL
for finite difference within their own inversion framework.

\subsection{Computational Fluid Dynamics}

Finally we describe three classical computational fluid dynamics examples to highlight the flexibility of Devito for another application domain. Additional CFD examples can be found in the Devito code repository in the form of a set of Jupyter notebooks. The three examples we describe here are the convection equation, the Burger equation and the Poisson  equation. These examples are adapted from~\citet{CFD12} and the example repository contains both the original Python implementation with Numpy and the implementation with Devito for comparison.

\subsubsection{Convection}

The convection governing equation for a field $u$ and a speed $c$ in two dimensions is:

\begin{equation}\label{eqn:convection}
\frac{\partial u}{\partial t}+c\frac{\partial u}{\partial x} + c\frac{\partial u}{\partial y} = 0.
\end{equation}

\begin{figure}
  \includegraphics[]{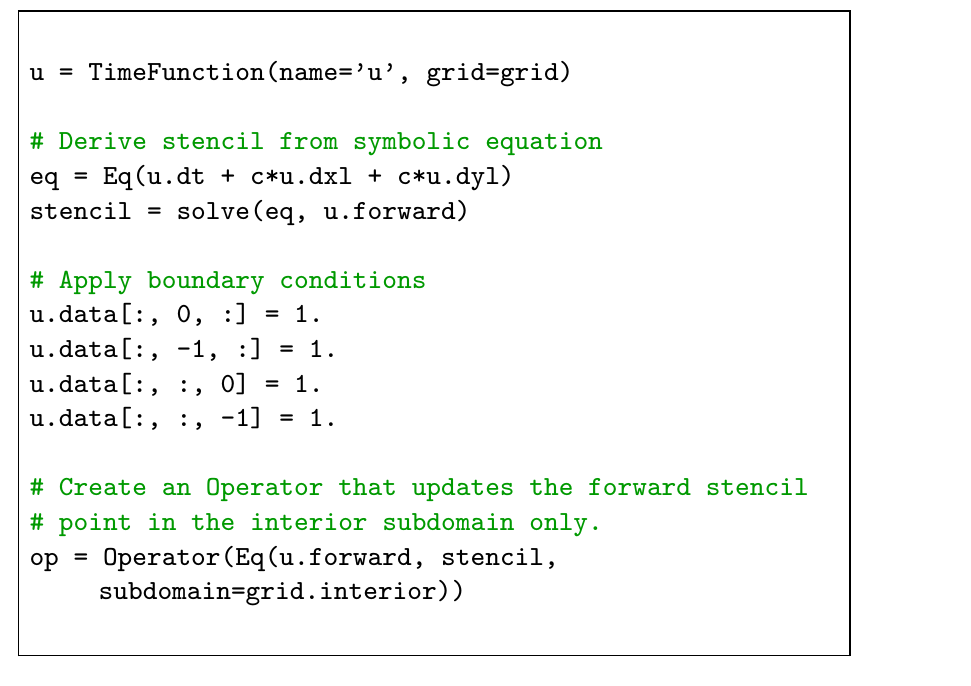}
\caption{Convection equation in Devito. In this example, the initial Dirichlet boundary conditions are set to $1$ using the API indexing feature, which allows to assign values to the \texttt{TensorFunction} data.}
\label{lst:convection}
\end{figure}

The same way we previously described it for the wave equation, $u$ is then defined as a \texttt{TimeFunction}.
In this simple case, the speed is a constant and does not need a symbolic representation, but a more general
definition of this equation is possible with the creation of $c$ as a Devito \texttt{Constant} that can accept any runtime value.
We then discretized the PDE using forward differences in time and backward differences in space:

\begin{equation}\label{eqn:convection_fd}
u_{i,j}^{n+1} = u_{i,j}^n-c \frac{\Delta t}{\Delta x}(u_{i,j}^n-u_{i-1,j}^n)-c \frac{\Delta t}{\Delta y}(u_{i,j}^n-u_{i,j-1}^n),
\end{equation}

which is implemented in Devito as in Fig.~\ref{lst:convection}.

\begin{figure}
   \begin{minipage}[t]{.45\textwidth}
	   \centering
   \includegraphics[width=\textwidth]{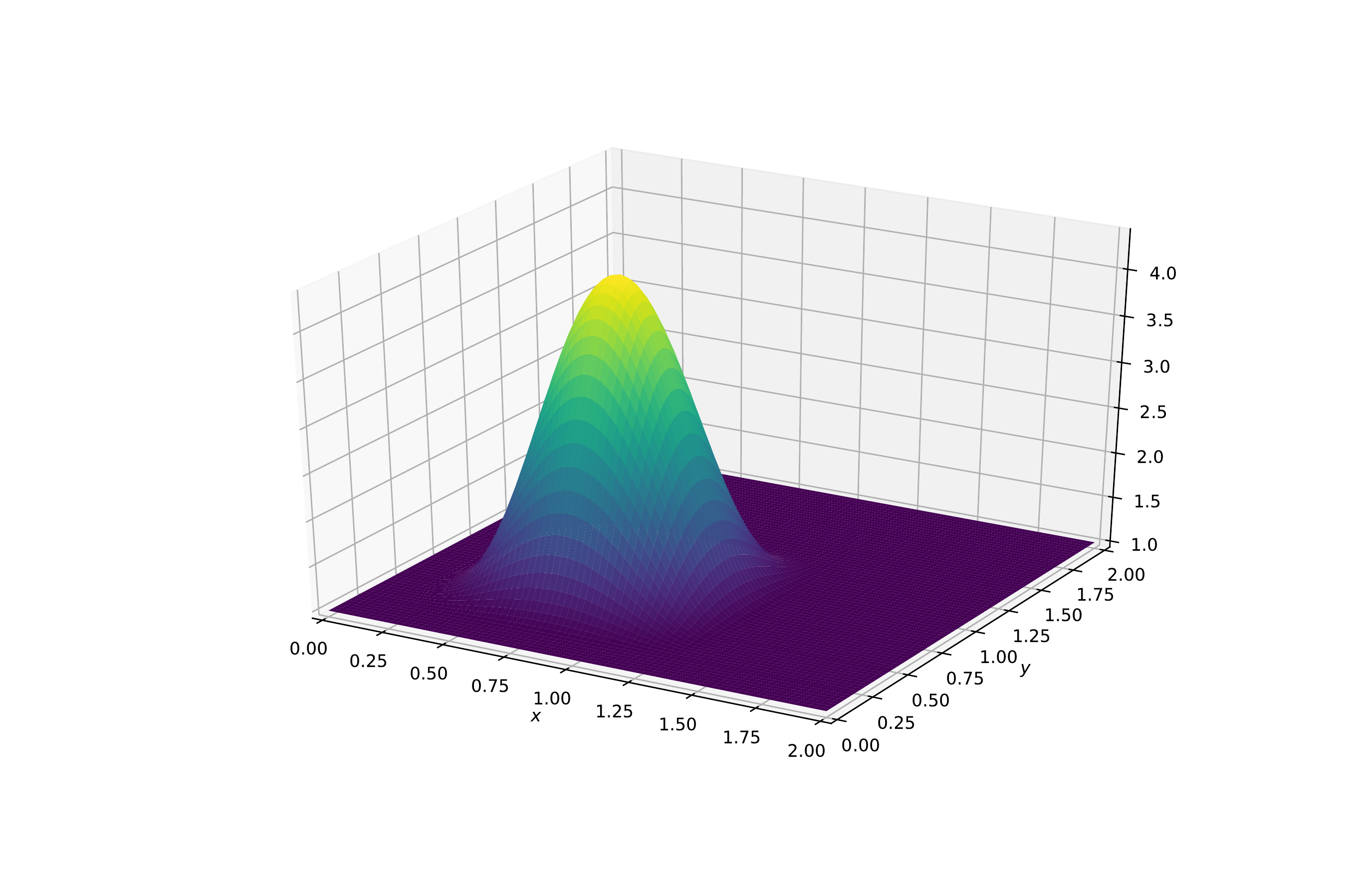}
   \end{minipage}
   \begin{minipage}[t]{.45\textwidth}
	   \centering
   \includegraphics[width=\textwidth]{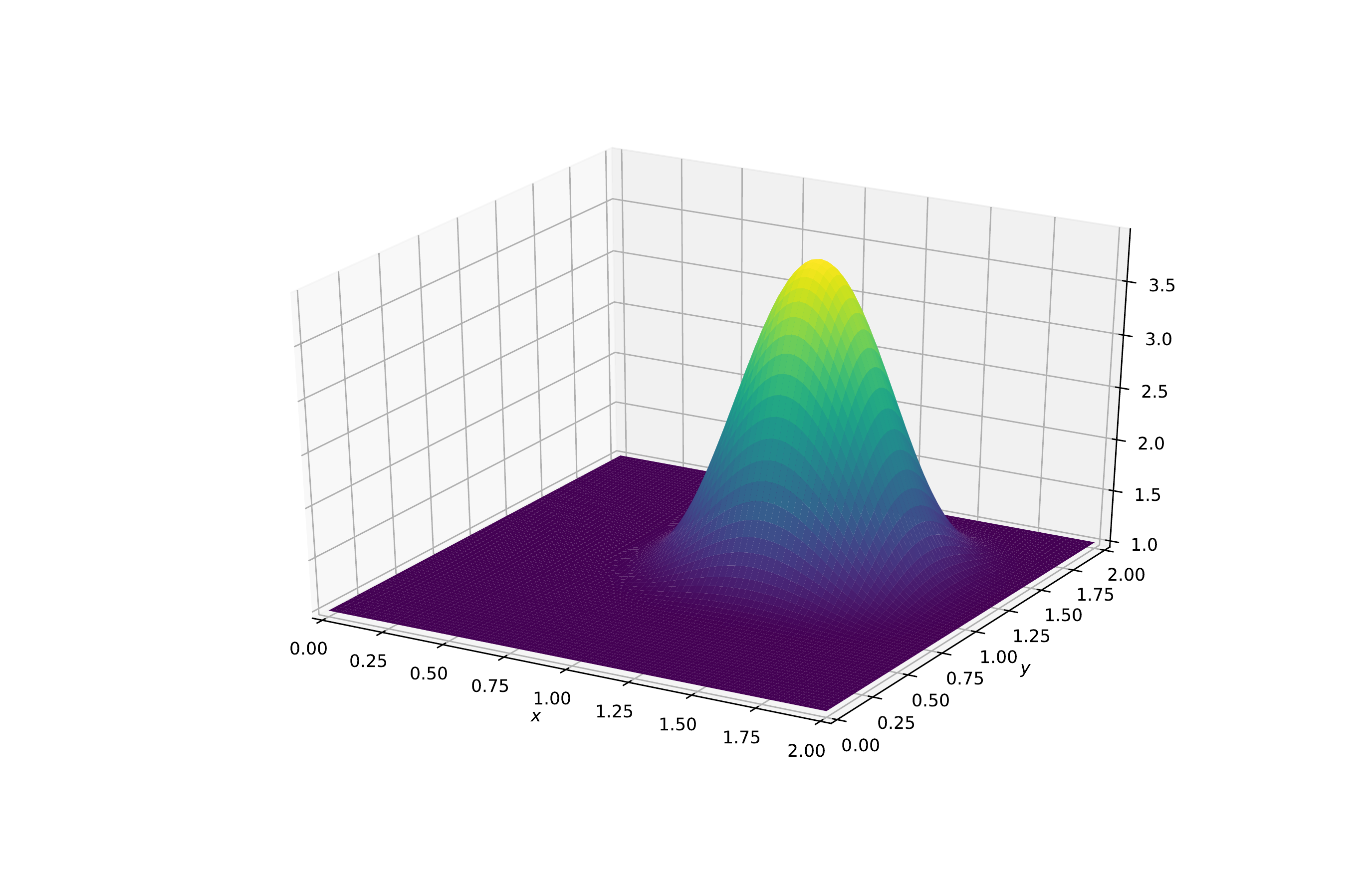}
   \end{minipage}
\caption{Initial (left) and final (right) time of the simulation of the convection equation.}
\label{fig:convection_res}
\end{figure}

The solution of the convection equation is displayed on Fig.~\ref{fig:convection_res} that shows the evolution of the field $u$ and the solution is consistent with the expected result produced by \citep{CFD12}.

\subsubsection{Burgers' equation}

\begin{figure}
  \includegraphics[]{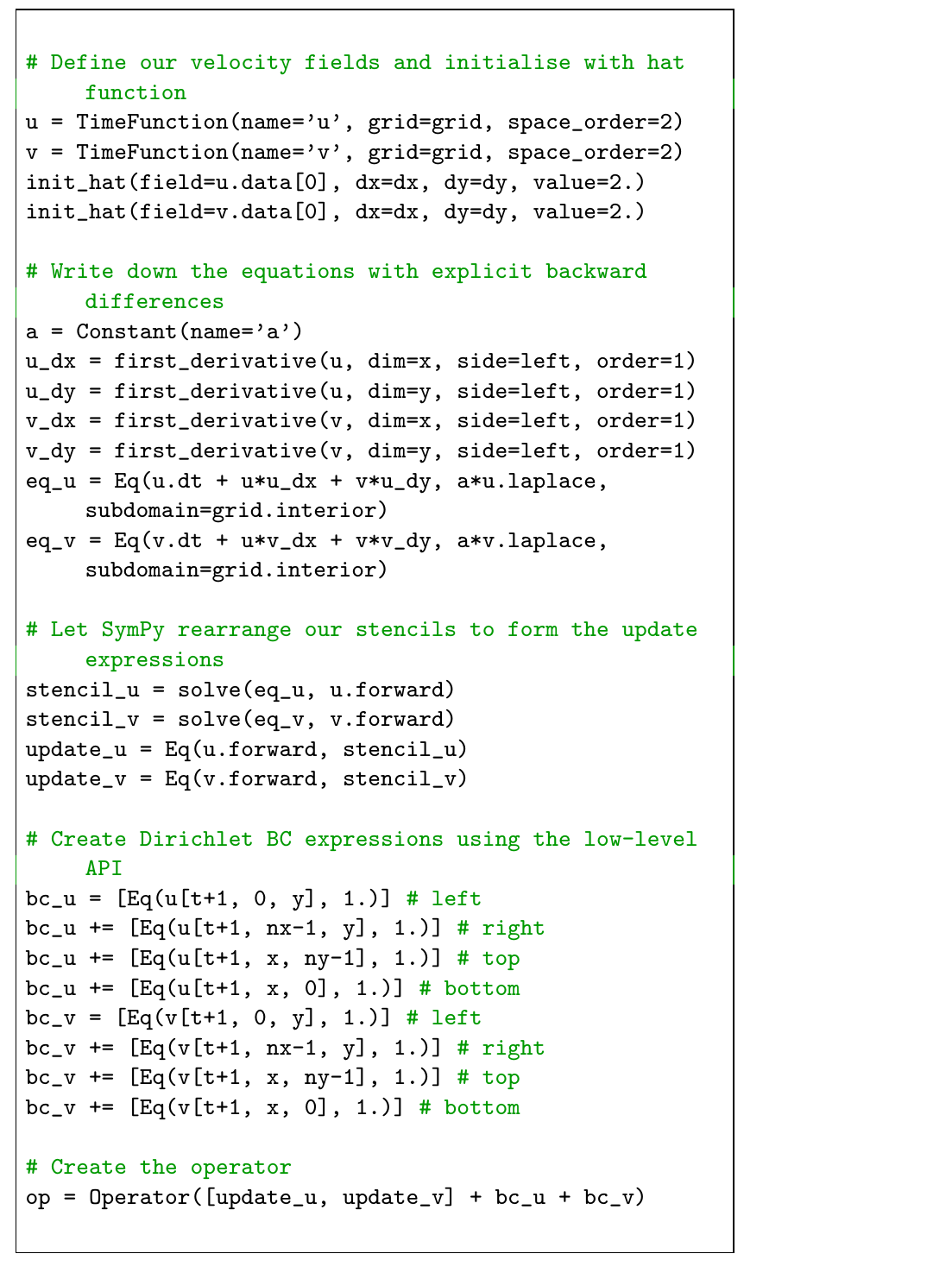}
\caption{Burgers' equations in Devito. {\color{black} In this example, we use explicitly the FD function \texttt{first\_derivative}. This function provides more flexibility and allows to take an upwind derivative, rather than a standard centered derivative (\texttt{\.dx}), to avoid odd-even coupling, which leads to chessboard artifacts in the solution.}}
\label{lst:burgers}
\end{figure}

In this second example, we show the solution of Burgers' equation. This example demonstrates that Devito supports coupled system of equations and non linear equations easily.
The Burgers' equation in two dimensions is defined as the following coupled PDE system:

\begin{align}\label{eqn:Burger}
\begin{cases}
 &\frac{\partial u}{\partial t} + u \frac{\partial u}{\partial x} + v \frac{\partial u}{\partial y} = \nu \; \left(\frac{\partial ^2 u}{\partial x^2} + \frac{\partial ^2 u}{\partial y^2}\right),\\
 &\frac{\partial v}{\partial t} + u \frac{\partial v}{\partial x} + v \frac{\partial v}{\partial y} = \nu \; \left(\frac{\partial ^2 v}{\partial x^2} + \frac{\partial ^2 v}{\partial y^2}\right),
\end{cases}
\end{align}

where $u, v$ are the two components of the solution and $\nu$ is the diffusion coefficient of the medium. The system of coupled equations is implemented in Devito in a few lines as shown in Fig.~\ref{lst:burgers}.

\begin{figure}
   \begin{minipage}[t]{.45\textwidth}
	   \centering
   \includegraphics[width=\textwidth]{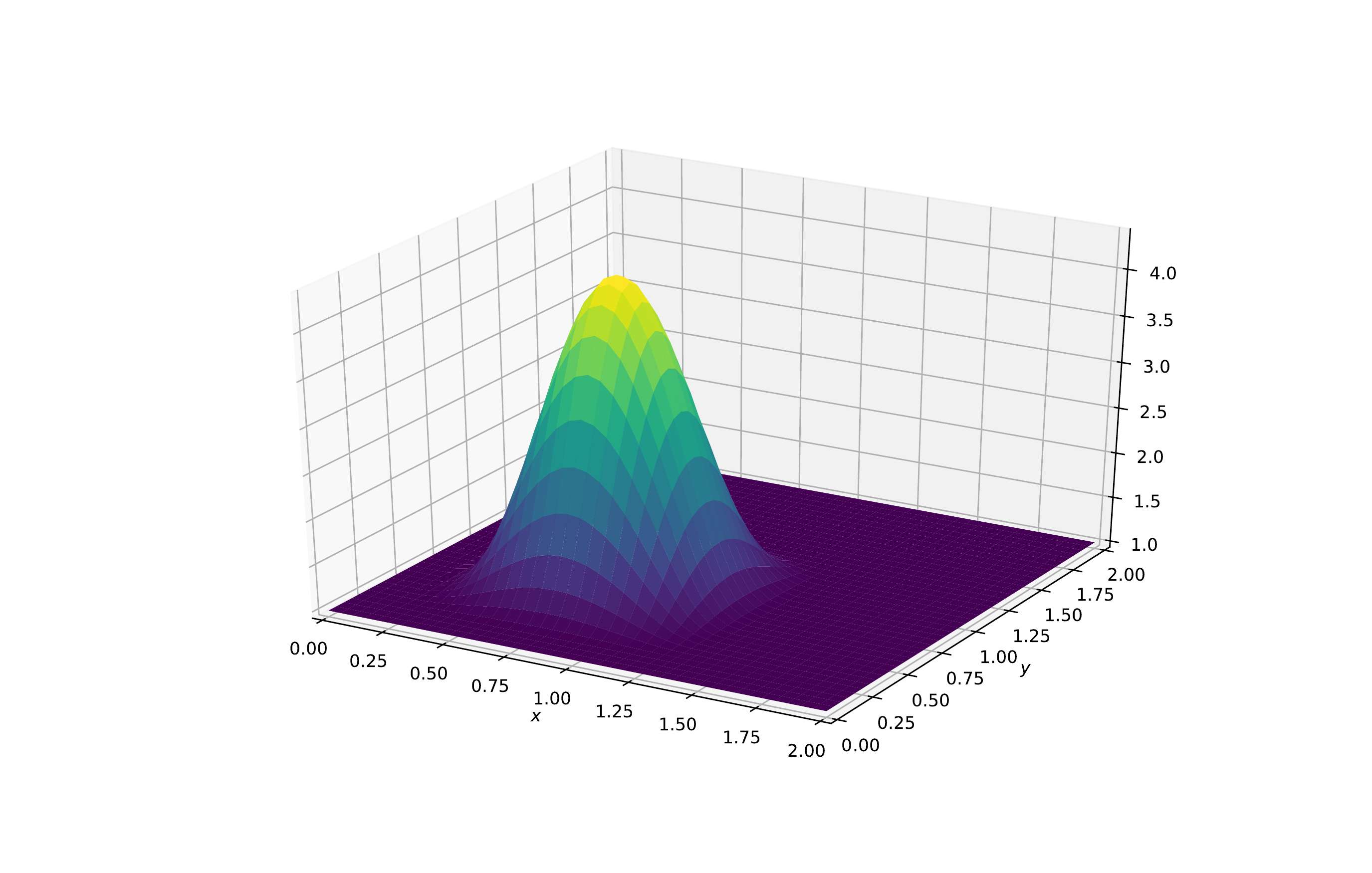}
   \end{minipage}
   \begin{minipage}[t]{.45\textwidth}
	   \centering
   \includegraphics[width=\textwidth]{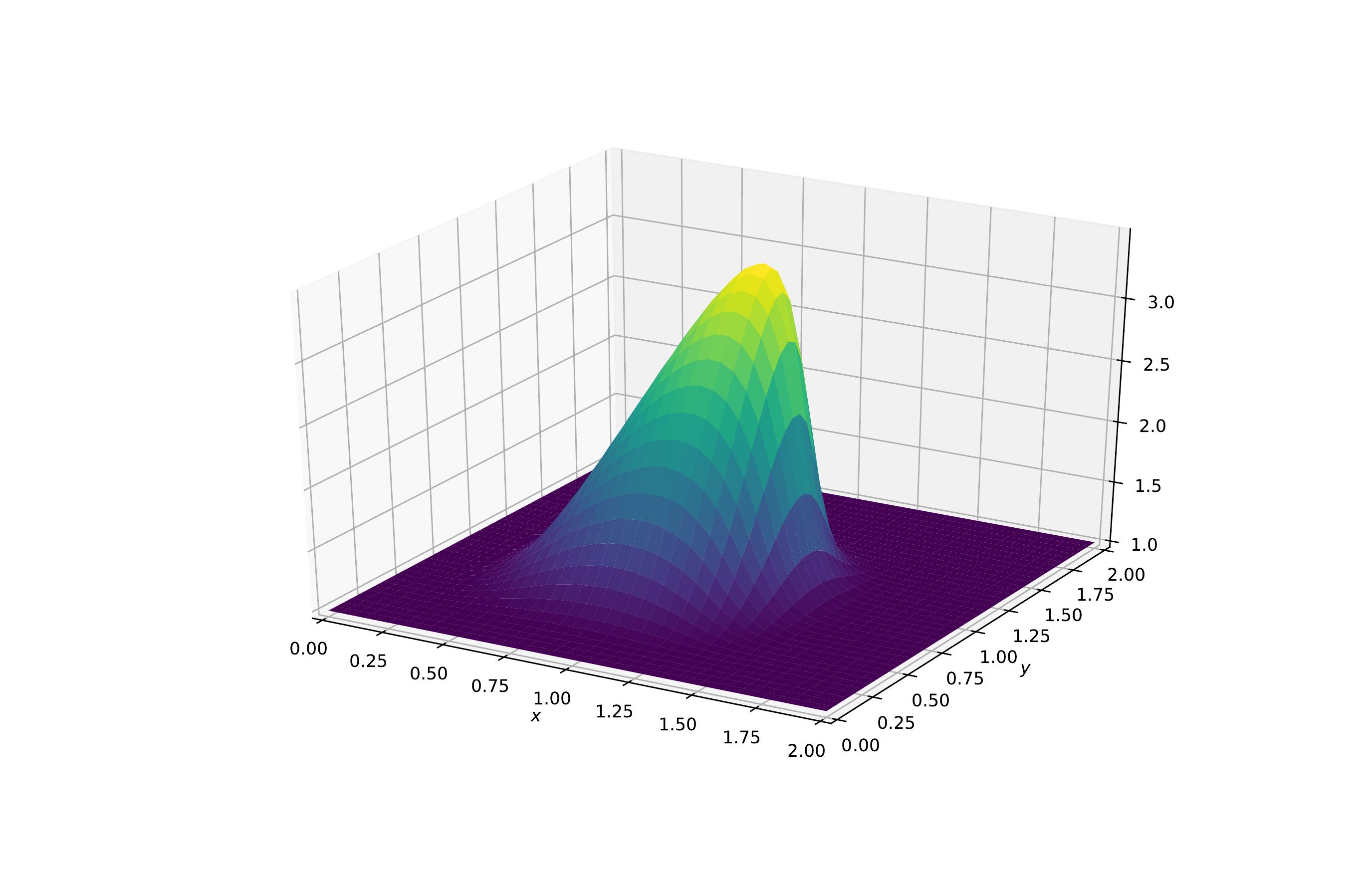}
   \end{minipage}
\caption{Initial (left) and final (right) time of the simulation of the Burgers' equations.}
\label{fig:burger_res}
\end{figure}

We show the initial state and the solution at the last time step of the Burgers' equation in Fig.~\ref{fig:burger_res}. Once again, the solution corresponds to the reference solution of \citet{CFD12}.

\subsubsection{Poisson}

\begin{figure} \centering
  \begin{minipage}[t][][t]{.49\textwidth}
    \includegraphics[scale=1.0]{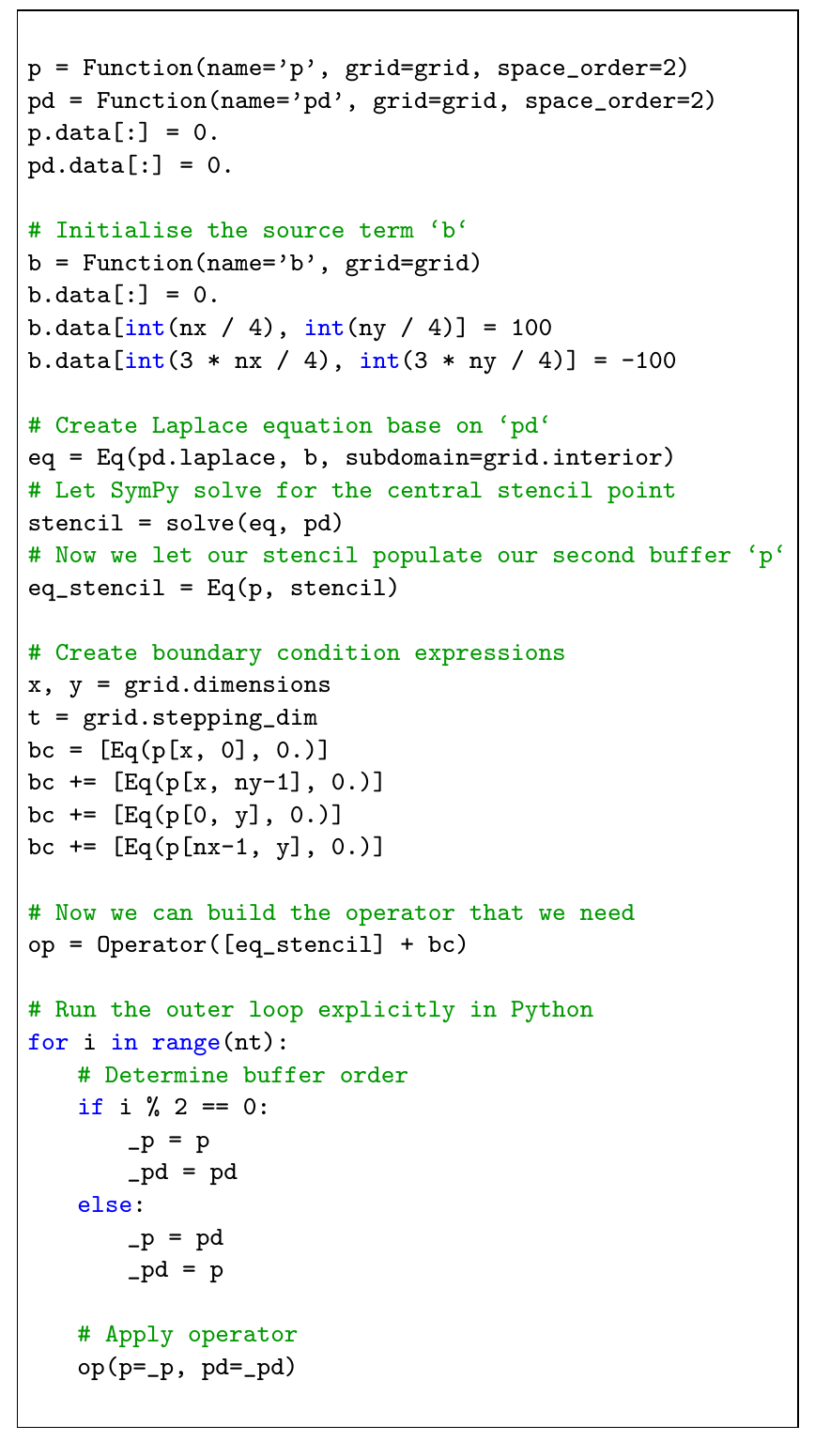}
    \caption{Poisson equation in Devito with field swap in Python.}
    \label{lst:poisson}
  \end{minipage}
  \begin{minipage}[t][][t]{.49\textwidth}
    \includegraphics[scale=1.0]{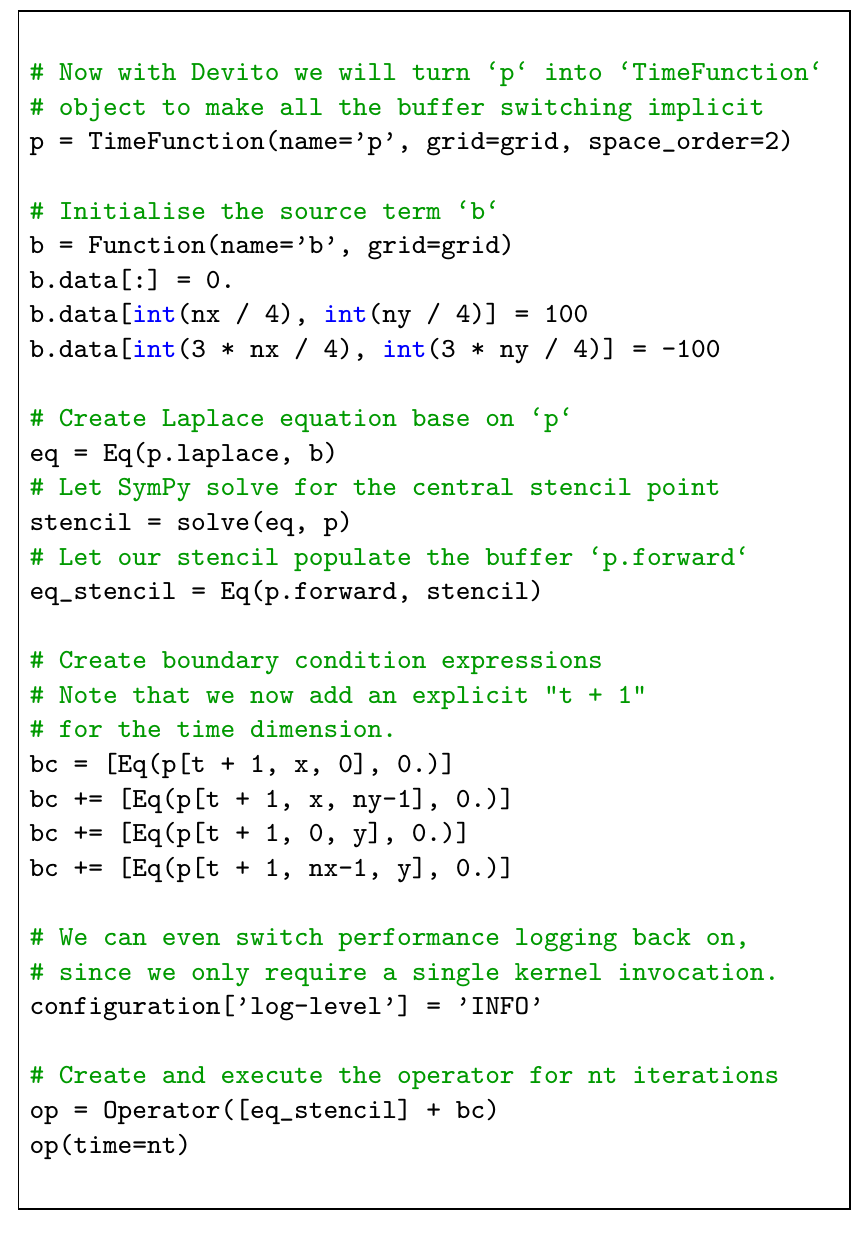}
    \caption{Poisson equation in Devito with buffered dimension for automatic swap at each iteration.}
    \label{lst:poisson_time}
  \end{minipage}
\end{figure}

We finally show the implementation of a solver for the Poisson equation in Devito. While the Poisson equation is not time dependent, the solution is obtained with an iterative solver and simplest one can easily be implemented with finite differences.
The Poisson equation for a field $p$ and a right hand side $b$ is defined as:

\begin{align}\label{eqn:Poisson}
 \frac{\partial ^2 p}{\partial x^2} + \frac{\partial ^2 p}{\partial y^2} = b,
\end{align}

and its solution can be computed iteratively with:

\begin{align}\label{eqn:Poisson_fd}
 p_{i,j}^{n+1}=\frac{(p_{i+1,j}^{n}+p_{i-1,j}^{n})\Delta y^2+(p_{i,j+1}^{n}+p_{i,j-1}^{n})\Delta x^2-b_{i,j}^{n}\Delta x^2\Delta y^2}{2(\Delta x^2+\Delta y^2)},
\end{align}

where the expression in Eq.~\ref{eqn:Poisson_fd} is computed until either the number of iterations is reached (our example case) or more realistically when $||p_{i,j}^{n+1} - p_{i,j}^{n}|| < \epsilon$.
We show two different implementations of a Poisson solver in Fig.~\ref{lst:poisson}, \ref{lst:poisson_time}. While these two implementations produce the same result, the second one takes advantage of Devito's \texttt{BufferedDimension} that allows to iterate automatically alternating between $p^{n}$ and $p^{n+1}$ as the two different time buffers in the \texttt{TimeFunction}.

\begin{figure}
   \begin{minipage}[t]{.45\textwidth}
	   \centering
   \includegraphics[width=\textwidth]{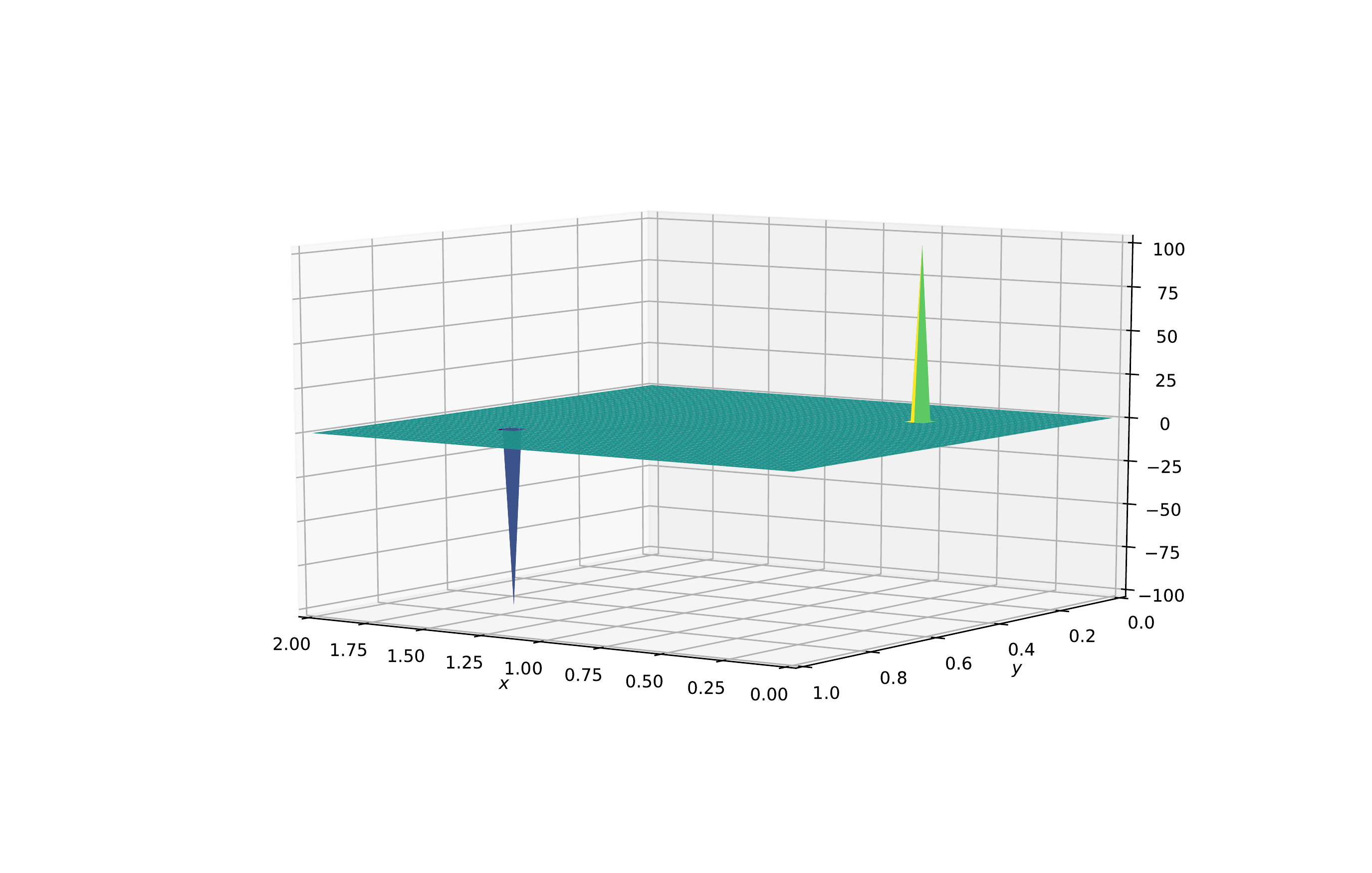}
   \end{minipage}
   \begin{minipage}[t]{.45\textwidth}
	   \centering
   \includegraphics[width=\textwidth]{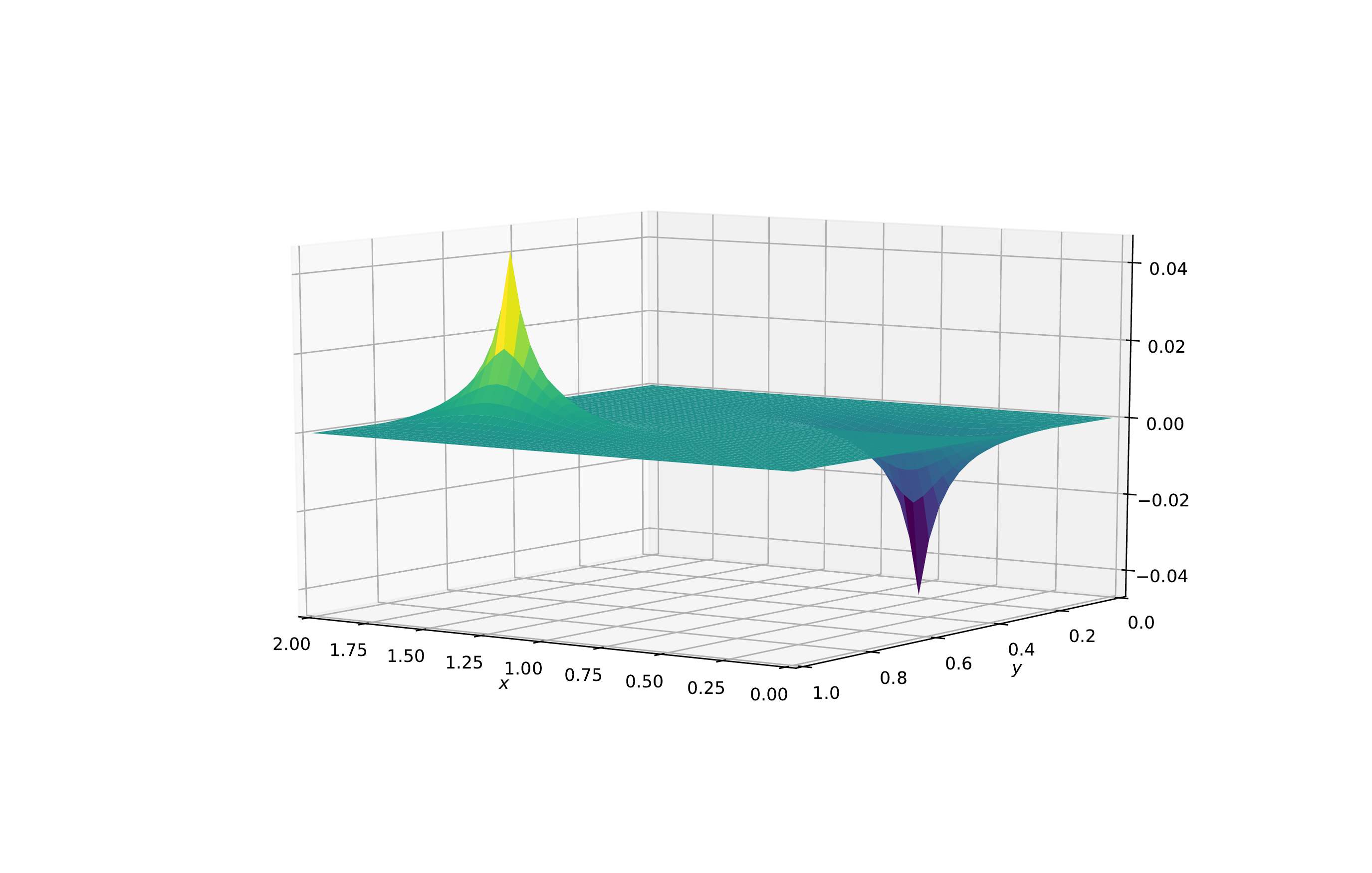}
   \end{minipage}
\caption{Right hand side (left) and solution (right) of the Poisson equations}
\label{fig:poisson_res}
\end{figure}

The solution of the Poisson equation is displayed on Fig.~\ref{fig:poisson_res} with its right-hand-side $b$.

These examples demonstrate the flexibility of Devito and show that a broad range of PDE can easily be implemented with Devito including non linear equation, coupled PDE system and steady state problems.

\section{Performance}\label{sec:performances}
In this section we demonstrate the performance of Devito from the
numerical and the inversion point of view, as well as the absolute performance
from the hardware point of view. This section only provides a brief overview of
Devito's performance and a more detailed description of the compiler and its
performance is covered in \citep{devito-compiler}.

\subsection{Error-cost analysis}
\begin{figure}
\centering
\includegraphics[width=0.8\linewidth]{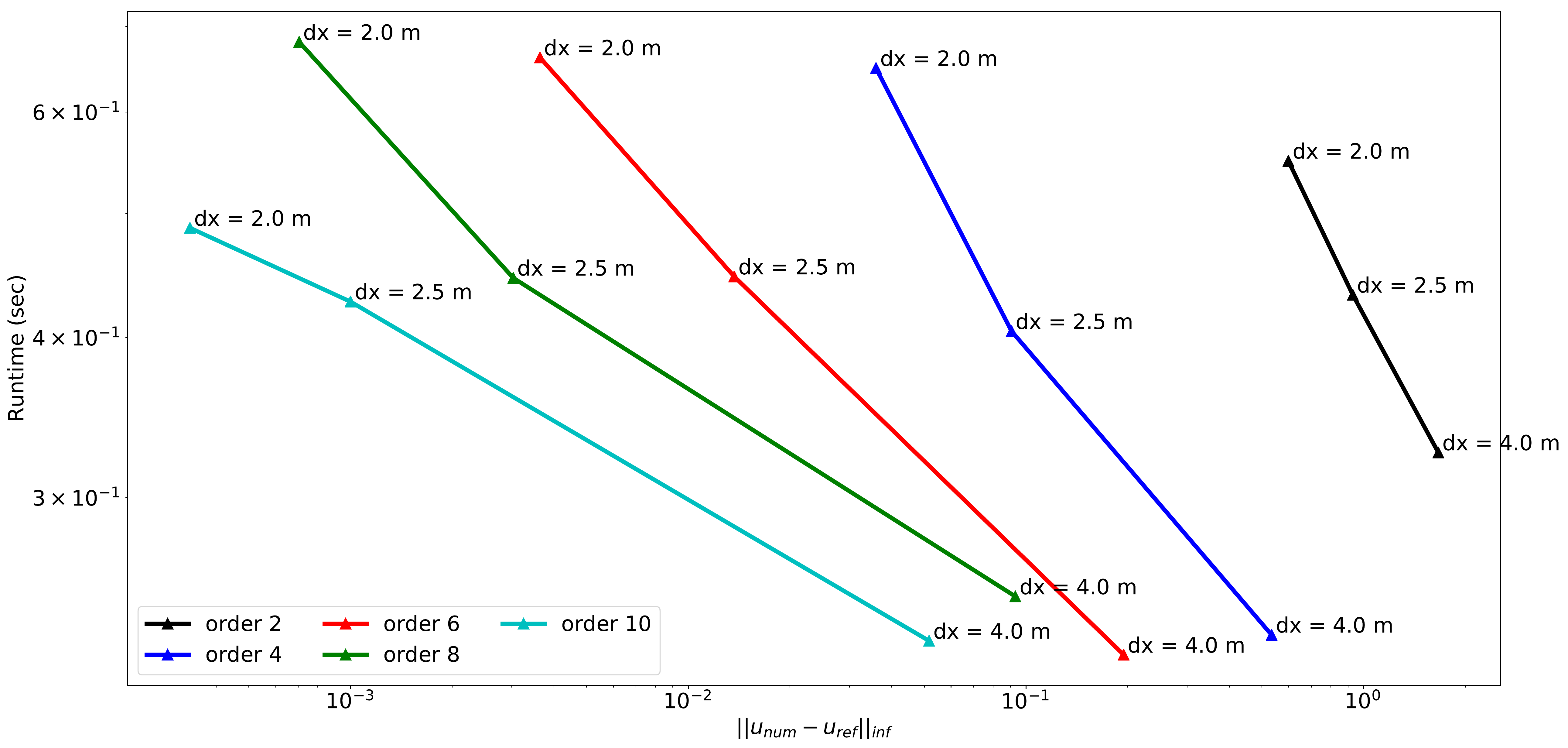}
\caption{Different spatial discretization orders accuracy against runtime
for a fixed physical setup (model size in $\unit{m}$ and propagation time).}
\label{fig:AvT}
\end{figure}

Devito's automatic code generation lets users define the spatial and temporal
order of FD stencils symbolically and without having to reimplement long
stencils by hand. This allows users to experiment with trade-offs between
discretization errors and runtime, as higher order FD stencils provide more
accurate solutions that come at increased runtime. For our error-cost
analysis, we compare absolute error in $L_2$-norm between the numerical and the
reference solution to the time-to-solution (the numerical and reference
solution are defined in the previous Sec.~\ref{sec:tests}).
Fig.~\ref{fig:AvT} shows the runtime and numerical error obtained for a fixed
physical setup. We use the same parameter as in Sec.~\ref{sec:convergence} with
a domain of $400 \unit{m} \times 400 \unit{m}$ and we simulate the wave
propagation for $150 \unit{ms}$.

The results in Fig.~\ref{fig:AvT} illustrate that higher order discretizations
produce a more accurate solution on a coarser grid with a smaller runtime. This
result is very useful for inverse problems, as a coarser grid requires less
memory and fewer time steps. A grid size two times bigger implies a reduction
of memory usage by a factor of $2^4$ for 3D modeling. Devito then allows users
to design FD simulators for inversion in an optimal way, where the
discretization order and grid size can be chosen according to the desired
numerical accuracy and availability of computational resources. {\color{black} While a near linear evolution of the runtime with increasing space order might be expected, we do not see such a behavior in practice. The main reason for this, is that the effect of Devito's performance optimizations for different space orders is difficult to predict and does not necessarily follow a linear relationship. On top of these optimizations, the runtimes also include the source injection and receiver interpolation, which impact the runtime in a non-linear way. Therefore these results are still acceptable.}. The order of
the FD stencils also affects the best possible hardware usage that can
theoretically be achieved and whether an algorithm is compute or memory bound,
a trade-off that is described by the roofline model.

\subsection{Roofline analysis}

We present performance results of our solver using the roofline model, as
previously discussed in~\citet{SevenDwarfs, asanovic2006landscape, Patterson,
williams2009roofline, louboutin2016ppf}. Given a finite difference scheme,
this method provides an estimate of the best achievable performance on the
underlying architecture, as well as an absolute measure of the hardware usage.
We also show a more classical metric, namely \emph{time to solution}, in
addition to the roofline plots, as both are essential for a clear picture of
the achieved performance. The experiments were run on an Intel Skylake 8180
architecture (28 physical cores, 38.5 MB shared L3 cache, with cores operating
at 2.5 Ghz). The observed Stream TRIAD \citep{McCalpin2007} was 105 GB/s. The
maximum single-precision FLOP performance was calculated as $\#\text{cores}
\cdot \# \text{avx units} \cdot \# \text{data items per vector register} \cdot
2 (\text{fused multiply-add)} \cdot \text{core frequency} = 4480 \
\unit{GFLOPs/s}$. A (more realistic) performance peak of $3285 \
\unit{GFLOPs/s}$ was determined by running the LINPACK benchmark
\citep{Dongarra}. These values are used to construct the roofline plots. In the performance results presented in this section, the operational intensity (OI) is computed by the Devito profiler from the symbolic expression after the compiler optimization. While the theoretical OI could be used, we chose to recompute it from the final optimized symbolic stencil for a more accurate performance measure. A more detailed overview of Devito's performance model is described in \citet{devito-compiler}.

\begin{figure}
\includegraphics[width=0.7\linewidth]{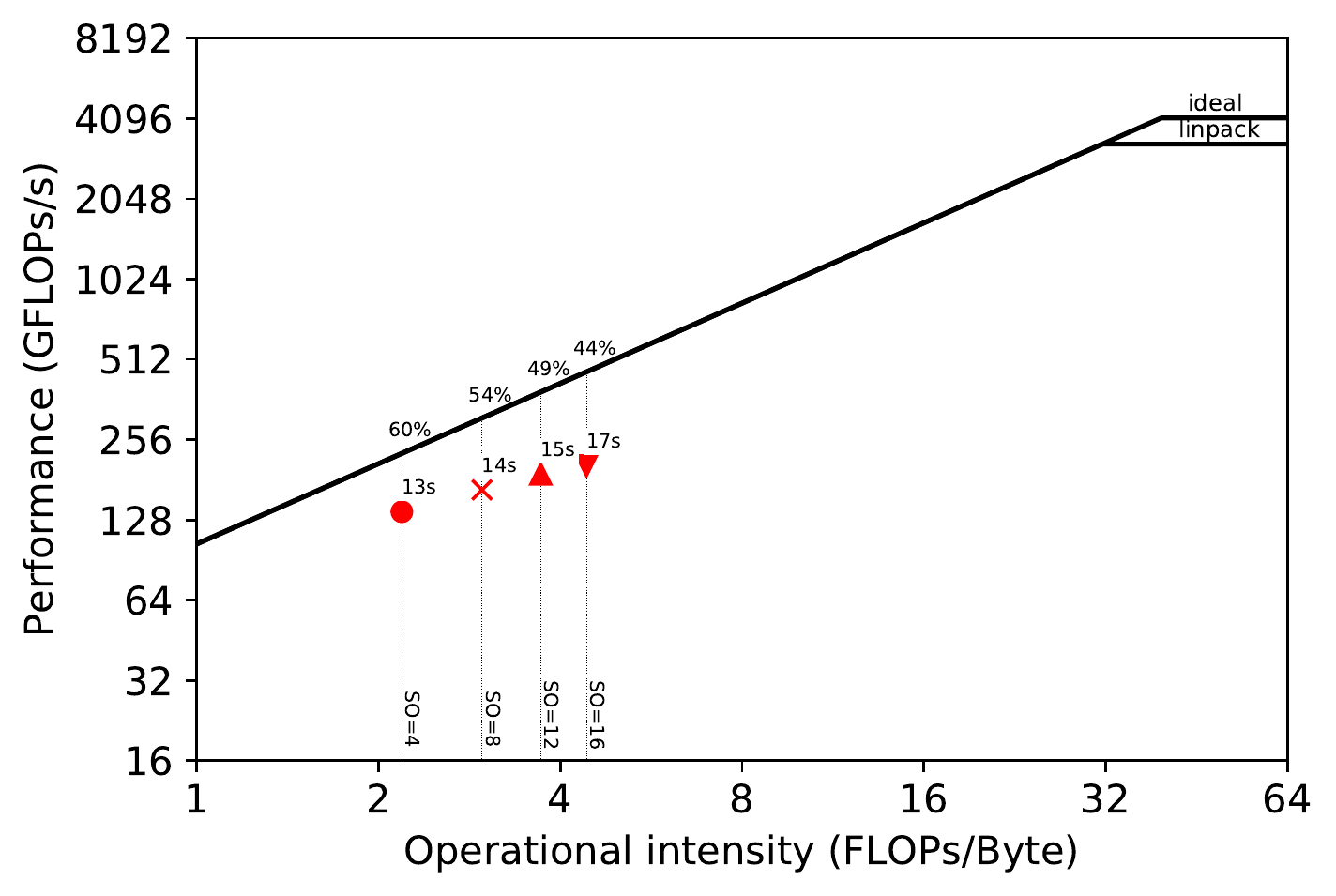}
\caption{Roofline plots for a $512\times512\times512$  model on a Skylake 8180
    architecture. The run times correspond to $1000\unit{ms}$ of modeling for
    four different spatial discretization orders (4, 8, 12, 16).}
\label{fig:roofline1}
\end{figure}

\begin{figure}
\includegraphics[width=0.7\linewidth]{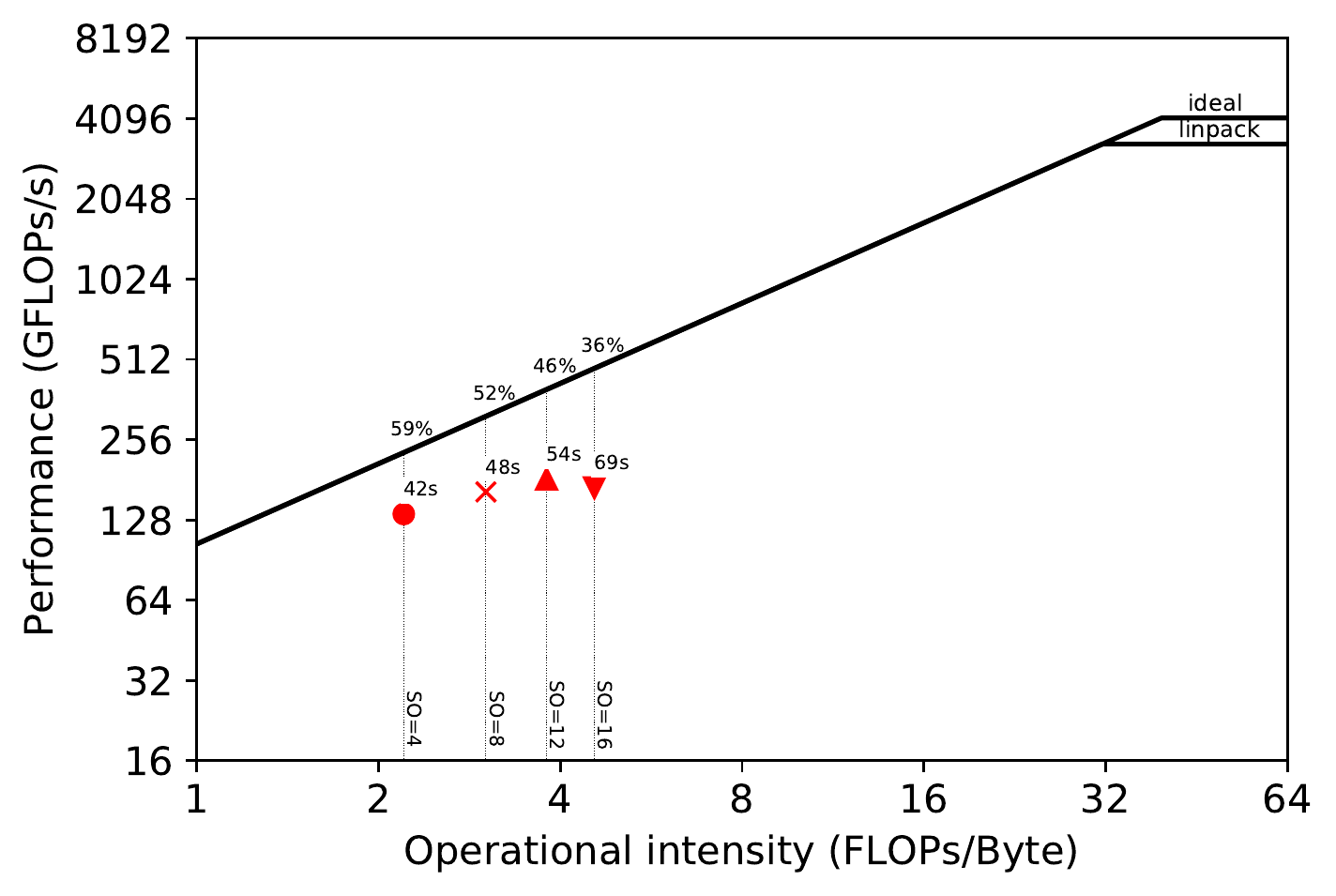}
\caption{Roofline plots for a $768\times768\times768$ model on a Skylake 8180
    architecture. The run times correspond to $1000\unit{ms}$ of modeling for
    four different spatial discretization orders (4, 8, 12, 16).}
\label{fig:roofline2}
\end{figure}

\begin{figure}
\includegraphics[width=0.7\linewidth]{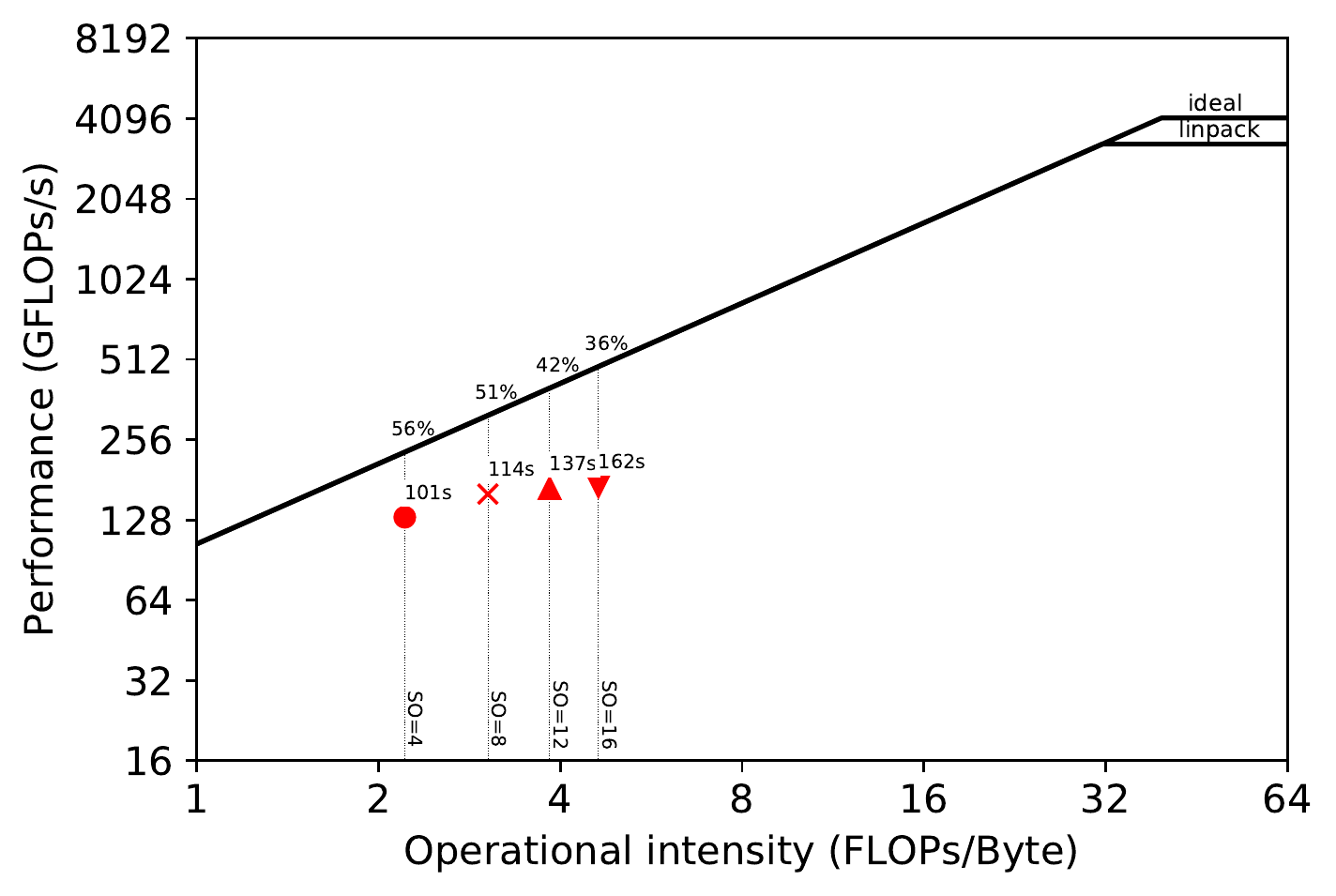}
\caption{Roofline plots for a $1024\times1024\times1024$  model on a Skylake
    8180 architecture. The run times correspond to $1000\unit{ms}$ of modeling
    for four different spatial discretization orders (4, 8, 12, 16).}
\label{fig:roofline3}
\end{figure}

We show three different roofline plots, one plot for each domain size
attempted, in Fig.~\ref{fig:roofline1}, ~\ref{fig:roofline2} and
~\ref{fig:roofline3}. Different space orders are represented as different data
points. The time-to-solution in seconds is annotated next to each data point.
The experiments were run with all performance optimizations enabled. Because
auto-tuning is used at runtime to determine the optimal loop-blocking
structure, timing only commences after autotuning has finished.  The reported
operational intensity benefits from the use of expression transformations as
described in Sec.~\ref{sec:devito}; particularly relevant for this problem is
the factorization of FD weights.

We observe that the time to solution increases nearly linearly with the size of
the domain. For example, for a 16th order discretization, we have a $17.1
\unit{sec}$ runtime for a $512\times512\times512$ domain and $162.6 \unit{sec}$
runtime for a $1024\times1024\times1024$ domain (8 times bigger domain and
about 9 times slower). This is not surprising: the computation lies in the
memory-bound regime and the working sets never fit in the L3 cache. We also
note a drop in performance with a 16th order discretization (relative to both
the other space orders and the attainable peak), especially when using larger
domains (Fig.~\ref{fig:roofline2} and \ref{fig:roofline3}). Our hypothesis,
supported by profiling with Intel VTune \citep{vtune}, is that this is due to
inefficient memory usage, in particular misaligned data accesses. Our plan to
improve the performance in this regime consists of resorting to a specialized
stencil optimizer such as YASK (see Sec.~\ref{sec:future-work}).  These
results show that we have a portable framework that achieves good performance
on different architectures. There is small room for improvements, as the
machine peak is still relatively distant, but ~50-60\% of the attainable peak
is usually considered very good. Finally, we remark that testing on new
architectures will only require extensions to the Devito compiler, if any,
while the application code remains unchanged.

\newpage
\section{Future Work}
\label{sec:future-work}
A key motivation for developing an embedded DSL such as Devito is to enable
quicker development, simpler maintenance, and better portability and performance of
solvers. The other benefit of this approach is that HPC developer effort can
be focused on developing the compiler technology that is reapplied to
a wide range of problems. This software reuse is fundamental to keep the
pace of technological evolution. For example, one of
the current projects in Devito regards the integration of YASK \citep{YASK},
a lower-level stencil optimizer conceived by Intel for Intel architectures.
Adding specialized backends such as YASK -- meaning that Devito can generate
and compile YASK code, rather than pure C/C++ -- is the key for long-term
performance portability, one of the goals that we are pursuing. Another motivation is to enable large-scale computations and as many different types of PDEs as possible. In practice, this means that a staggered grid setup with half-node discretization and domain decomposition will be required. These two main requirements to extend the DSL to a broader community and applications are in full development and will be made available in future releases.

\conclusions  

We have introduced a DSL for time-domain simulation for inversion and its
application to a seismic inverse problem based on the finite difference method.
Using the Devito DSL a highly optimized and parallel finite difference solver
can be implemented within just a few lines of Python code. Although the current
application focuses on features required for seismic imaging applications,
Devito can already be used in problems based on other equations; a series of
CFD examples are included in the code repository.

The code traditionally used to solve such problems is highly complex. The
primary reason for this is that the complexity
introduced by the mathematics is interleaved with the complexity introduced by
performance engineering of the code to make it useful for practical use. By
introducing a separation of concerns, these aspects are decoupled and both
simplified. Devito successfully achieves this decoupling while delivering good
computational performance and maintaining generality, both of which shall
continue to be improved in future versions.

\section{Code Availability}

The code source code, examples and test script are available on github at
\url{https://github.com/opesci/devito} and contains a README for installation.
A more detailed overview of the project, list of publication and documentation
of the software generated with Sphinx is available at
\url{http://www.devitoproject.org/}. To install Devito:

\begin{verbatim}
	git clone -b v3.1.0 https://github.com/opesci/devito
	cd devito
	conda env create -f environment.yml
	source activate devito
	pip install -e .
\end{verbatim}


\begin{acknowledgements}
The development of Devito was primarily supported through the Imperial
College London Intel Parallel Computing Centre. We would also like to acknowledgement
the support from the SINBAD II project and support of the member organizations
of the SINBAD Consortium as well as EPSRC grants EP/M011054/1 and EP/L000407/1.
\end{acknowledgements}



\clearpage
\bibliographystyle{copernicus}
\bibliography{bib_GMD}

\begin{thebibliography}{75}
\providecommand{\natexlab}[1]{#1}
\providecommand{\url}[1]{{\tt #1}}
\providecommand{\urlprefix}{URL }
\expandafter\ifx\csname urlstyle\endcsname\relax
  \providecommand{\doi}[1]{doi:\discretionary{}{}{}#1}\else
  \providecommand{\doi}{doi:\discretionary{}{}{}\begingroup
  \urlstyle{rm}\Url}\fi

\bibitem[{Aln{\ae}s et~al.(2014)Aln{\ae}s, Logg, {\O}lgaard, Rognes, and
  Wells}]{Alnaes2014}
Aln{\ae}s, M.~S., Logg, A., {\O}lgaard, K.~B., Rognes, M.~E., and Wells, G.~N.:
  {U}nified {F}orm {L}anguage: a domain-specific language for weak formulations
  of partial differential equations, ACM Transactions on Mathematical Software
  (TOMS), 40, 9, 2014.

\bibitem[{Andreolli et~al.(2015)Andreolli, Thierry, Borges, Skinner, and
  Yount}]{Andreolli2015}
Andreolli, C., Thierry, P., Borges, L., Skinner, G., and Yount, C.: Chapter 23
  - Characterization and Optimization Methodology Applied to Stencil
  Computations, in: High Performance Parallelism Pearls, edited by Reinders, J.
  and Jeffers, J., pp. 377 -- 396, Morgan Kaufmann, Boston,
  \doi{http://dx.doi.org/10.1016/B978-0-12-802118-7.00023-6},
  \urlprefix\url{http://www.sciencedirect.com/science/article/pii/B9780128021187000236},
  2015.

\bibitem[{Arbona et~al.(2017)Arbona, Mi{\~n}ano, Rigo, Bona, Palenzuela,
  Artigues, Bona-Casas, and Mass{\'o}}]{simflowny}
Arbona, A., Mi{\~n}ano, B., Rigo, A., Bona, C., Palenzuela, C., Artigues, A.,
  Bona-Casas, C., and Mass{\'o}, J.: Simflowny 2: An upgraded platform for
  scientific modeling and simulation, arXiv preprint arXiv:1702.04715, 2017.

\bibitem[{Asanovic et~al.(2006)Asanovic, Bodik, Catanzaro, Gebis, Husbands,
  Keutzer, Patterson, Plishker, Shalf, Williams et~al.}]{asanovic2006landscape}
Asanovic, K., Bodik, R., Catanzaro, B.~C., Gebis, J.~J., Husbands, P., Keutzer,
  K., Patterson, D.~A., Plishker, W.~L., Shalf, J., Williams, S.~W., et~al.:
  The landscape of parallel computing research: A view from berkeley, Tech.
  rep., Technical Report UCB/EECS-2006-183, EECS Department, University of
  California, Berkeley, 2006.

\bibitem[{Backus(1978)}]{HistoryOfFortran}
Backus, J.: The history of Fortran I, II, and III, in: History of programming
  languages I, pp. 25--74, ACM, 1978.

\bibitem[{Barba and Forsyth(2018)}]{CFD12}
Barba, L.~A. and Forsyth, G.~F.: CFD Python: the 12 steps to Navier-Stokes
  equations., \doi{https://doi.org/10.21105/jose.00021}, 2018.

\bibitem[{Bondhugula et~al.(2008)Bondhugula, Hartono, Ramanujam, and
  Sadayappan}]{pluto}
Bondhugula, U., Hartono, A., Ramanujam, J., and Sadayappan, P.: A Practical
  Automatic Polyhedral Parallelizer and Locality Optimizer, in: Proceedings of
  the 2008 ACM SIGPLAN Conference on Programming Language Design and
  Implementation, PLDI '08, pp. 101--113, ACM, New York, NY, USA,
  \doi{10.1145/1375581.1375595},
  \urlprefix\url{http://doi.acm.org/10.1145/1375581.1375595}, 2008.

\bibitem[{C{\'a}rdenas and Karplus(1970)}]{PDEL}
C{\'a}rdenas, A.~F. and Karplus, W.~J.: PDEL---a language for partial
  differential equations, Communications of the ACM, 13, 184--191, 1970.

\bibitem[{Cauchy(1847)}]{Cauchy1847Methode}
Cauchy, A.-L.: {M\'{e}thode g\'{e}n\'{e}rale pour la r\'{e}solution des
  syst\`{e}mes d'\'{e}quations simultan\'{e}es}, Compte Rendu des S\'{e}ances
  de L'Acad\'{e}mie des Sciences XXV, S\'{e}rie A, 536--538, 1847.

\bibitem[{Christen et~al.(2011)Christen, Schenk, and Burkhart}]{PATUS}
Christen, M., Schenk, O., and Burkhart, H.: PATUS: A Code Generation and
  Autotuning Framework for Parallel Iterative Stencil Computations on Modern
  Microarchitectures, in: Proceedings of the 2011 IEEE International Parallel
  \& Distributed Processing Symposium, IPDPS '11, pp. 676--687, IEEE Computer
  Society, Washington, DC, USA, \doi{10.1109/IPDPS.2011.70},
  \urlprefix\url{http://dx.doi.org/10.1109/IPDPS.2011.70}, 2011.

\bibitem[{Clayton and Engquist(1977)}]{Clayton1529}
Clayton, R. and Engquist, B.: Absorbing boundary conditions for acoustic and
  elastic wave equations, Bulletin of the Seismological Society of America, 67,
  1529--1540,
  \urlprefix\url{http://bssa.geoscienceworld.org/content/67/6/1529}, 1977.

\bibitem[{Colella(2004)}]{SevenDwarfs}
Colella, P.: Defining Software Requirements for Scientific Computing, 2004.

\bibitem[{Cook~Jr(1988)}]{alpal}
Cook~Jr, G.~O.: ALPAL: A tool for the development of large-scale simulation
  codes, Tech. rep., Lawrence Livermore National Lab., CA (USA), 1988.

\bibitem[{Dongarra(1988)}]{Dongarra}
Dongarra, J.: The {LINPACK} Benchmark: {An} Explanation, in: Proceedings of the
  1st International Conference on Supercomputing, pp. 456--474,
  Springer-Verlag, London, UK, UK,
  \urlprefix\url{http://dl.acm.org/citation.cfm?id=647970.742568}, 1988.

\bibitem[{Farrell et~al.(2013)Farrell, Ham, Funke, and Rognes}]{DolfinAdjoint}
Farrell, P.~E., Ham, D.~A., Funke, S.~W., and Rognes, M.~E.: Automated
  Derivation of the Adjoint of High-Level Transient Finite Element Programs,
  SIAM Journal on Scientific Computing, 35, C369--C393,
  \doi{10.1137/120873558}, \urlprefix\url{http://dx.doi.org/10.1137/120873558},
  2013.

\bibitem[{{Fomel et. al}(2013)}]{Madagascar}
{Fomel et. al}, S.: Madagascar: open-source software project for
  multidimensional data analysis and reproducible computational experiments,
  \doi{http://doi.org/10.5334/jors.ag}, 2013.

\bibitem[{Griewank and Walther(2000)}]{Griewank:2000:ARI:347837.347846}
Griewank, A. and Walther, A.: Algorithm 799: Revolve: An Implementation of
  Checkpointing for the Reverse or Adjoint Mode of Computational
  Differentiation, ACM Trans. Math. Softw., 26, 19--45,
  \doi{10.1145/347837.347846},
  \urlprefix\url{http://doi.acm.org/10.1145/347837.347846}, 2000.

\bibitem[{Haber et~al.(2012)Haber, Chung, and Herrmann}]{haber10TRemp}
Haber, E., Chung, M., and Herrmann, F.~J.: An effective method for parameter
  estimation with {PDE} constraints with multiple right hand sides, SIAM
  Journal on Optimization, 22,
  \urlprefix\url{http://dx.doi.org/10.1137/11081126X}, 2012.

\bibitem[{Hawick and Playne(2013)}]{STARGATES}
Hawick, K.~A. and Playne, D.~P.: Simulation Software Generation using a
  Domain-Specific Language for Partial Differential Field Equations, in: 11th
  International Conference on Software Engineering Research and Practice
  (SERP'13), CSTN-187, p. SER3829, WorldComp, Las Vegas, USA, 2013.

\bibitem[{Henretty et~al.(2013)Henretty, Veras, Franchetti, Pouchet, Ramanujam,
  and Sadayappan}]{StencilDSL-1}
Henretty, T., Veras, R., Franchetti, F., Pouchet, L.-N., Ramanujam, J., and
  Sadayappan, P.: A Stencil Compiler for Short-vector SIMD Architectures, in:
  Proceedings of the 27th International ACM Conference on International
  Conference on Supercomputing, ICS '13, pp. 13--24, ACM, New York, NY, USA,
  \doi{10.1145/2464996.2467268},
  \urlprefix\url{http://doi.acm.org/10.1145/2464996.2467268}, 2013.

\bibitem[{Hopper(1952)}]{hopperA0}
Hopper, G.~M.: The education of a computer, in: Proceedings of the 1952 ACM
  national meeting (Pittsburgh), pp. 243--249, ACM, 1952.

\bibitem[{Igel(2016)}]{id2138}
Igel, H.: {Computational Seismology: A Practical Introduction}, Oxford
  University Press, 1. edn.,
  \urlprefix\url{https://global.oup.com/academic/product/computational-seismology-9780198717409?cc=de{\&}lang=en{\&}},
  2016.

\bibitem[{{Intel Corporation}(2016)}]{vtune}
{Intel Corporation}: {Intel VTune Performance Analyzer},
  https://software.intel.com/en-us/intel-vtune-amplifier-xe, 2016.

\bibitem[{Iverson(1962)}]{Iverson1962}
Iverson, K.~E.: A Programming Language, John Wiley \& Sons, Inc., New York, NY,
  USA, 1962.

\bibitem[{Jacobs et~al.(2016)Jacobs, Jammy, and Sandham}]{JacobsSBLI}
Jacobs, C.~T., Jammy, S.~P., and Sandham, N.~D.: OpenSBLI: {A} framework for
  the automated derivation and parallel execution of finite difference solvers
  on a range of computer architectures, CoRR, abs/1609.01277,
  \urlprefix\url{http://arxiv.org/abs/1609.01277}, 2016.

\bibitem[{Jones(1954)}]{jones1954survey}
Jones, J.~L.: A survey of automatic coding techniques for digital computers,
  Ph.D. thesis, Massachusetts Institute of Technology, 1954.

\bibitem[{K{\"o}ster et~al.(2014)K{\"o}ster, Lei{\ss}a, Hack, Membarth, and
  Slusallek}]{koster2014platform}
K{\"o}ster, M., Lei{\ss}a, R., Hack, S., Membarth, R., and Slusallek, P.:
  Platform-Specific Optimization and Mapping of Stencil Codes through
  Refinement, in: Proceedings of the 1st International Workshop on
  High-Performance Stencil Computations, pp. 1--6, 2014.

\bibitem[{Kukreja et~al.(2018)Kukreja, H{\"u}ckelheim, Lange, Louboutin,
  Walther, Funke, and Gorman}]{kukreja2018high}
Kukreja, N., H{\"u}ckelheim, J., Lange, M., Louboutin, M., Walther, A., Funke,
  S.~W., and Gorman, G.: High-level python abstractions for optimal
  checkpointing in inversion problems, arXiv preprint arXiv:1802.02474, 2018.

\bibitem[{Kumar et~al.(2015)Kumar, Wason, and Herrmann}]{kumar2015sss}
Kumar, R., Wason, H., and Herrmann, F.~J.: Source separation for simultaneous
  towed-streamer marine acquisition {\textendash}- a compressed sensing
  approach, Geophysics, 80, WD73--WD88, \doi{10.1190/geo2015-0108.1},
  \urlprefix\url{https://www.slim.eos.ubc.ca/Publications/Public/Journals/Geophysics/2015/kumar2015sss/kumar2015sss_revised.pdf},
  (Geophysics), 2015.

\bibitem[{{L}ange et~al.(2017){L}ange, {K}ukreja, {L}uporini, {L}ouboutin,
  {Y}ount, {H}\"uckelheim, and {G}orman}]{Lange2017scipy}
{L}ange, M., {K}ukreja, N., {L}uporini, F., {L}ouboutin, M., {Y}ount, C.,
  {H}\"uckelheim, J., and {G}orman, G.~J.: {O}ptimised finite difference
  computation from symbolic equations, in: {P}roceedings of the 15th {P}ython
  in {S}cience {C}onference, edited by {K}aty {H}uff, {D}avid {L}ippa, {D}illon
  {N}iederhut, and {P}acer, M., pp. 89 -- 96, 2017.

\bibitem[{Lions(1971)}]{LionsJL1971}
Lions, J.~L.: Optimal control of systems governed by partial differential
  equations, Springer-Verlag Berlin Heidelberg, 1st edn., 1971.

\bibitem[{Liu and Fomel(2011)}]{doi:10.1190/geo2010-0231.1}
Liu, Y. and Fomel, S.: Seismic data interpolation beyond aliasing using
  regularized nonstationary autoregression, GEOPHYSICS, 76, V69--V77,
  \doi{10.1190/geo2010-0231.1},
  \urlprefix\url{https://doi.org/10.1190/geo2010-0231.1}, 2011.

\bibitem[{Logg et~al.(2012)Logg, Mardal, Wells et~al.}]{Fenics}
Logg, A., Mardal, K.-A., Wells, G.~N., et~al.: Automated Solution of
  Differential Equations by the Finite Element Method, Springer,
  \doi{10.1007/978-3-642-23099-8}, 2012.

\bibitem[{Louboutin et~al.(2017{\natexlab{a}})Louboutin, Lange, Herrmann,
  Kukreja, and Gorman}]{louboutin2016ppf}
Louboutin, M., Lange, M., Herrmann, F.~J., Kukreja, N., and Gorman, G.:
  Performance prediction of finite-difference solvers for different computer
  architectures, Computers \& Geosciences, 105, 148--157,
  \doi{https://doi.org/10.1016/j.cageo.2017.04.014}, 2017{\natexlab{a}}.

\bibitem[{Louboutin et~al.(2017{\natexlab{b}})Louboutin, Witte, Lange, Kukreja,
  Luporini, Gorman, and Herrmann}]{TLE1}
Louboutin, M., Witte, P., Lange, M., Kukreja, N., Luporini, F., Gorman, G., and
  Herrmann, F.~J.: Full-waveform inversion, Part 1: Forward modeling, The
  Leading Edge, 36, 1033--1036, \doi{10.1190/tle36121033.1},
  \urlprefix\url{https://doi.org/10.1190/tle36121033.1}, 2017{\natexlab{b}}.

\bibitem[{Luporini et~al.(2015)Luporini, Varbanescu, Rathgeber, Bercea,
  Ramanujam, Ham, and Kelly}]{Luporini2015}
Luporini, F., Varbanescu, A.~L., Rathgeber, F., Bercea, G.-T., Ramanujam, J.,
  Ham, D.~A., and Kelly, P. H.~J.: Cross-Loop Optimization of Arithmetic
  Intensity for Finite Element Local Assembly, ACM Trans. Archit. Code Optim.,
  11, 57:1--57:25, \doi{10.1145/2687415},
  \urlprefix\url{http://doi.acm.org/10.1145/2687415}, 2015.

\bibitem[{{Luporini} et~al.(2018){Luporini}, {Lange}, {Louboutin}, {Kukreja},
  {H{\"u}ckelheim}, {Yount}, {Witte}, {Kelly}, {Gorman}, and
  {Herrmann}}]{devito-compiler}
{Luporini}, F., {Lange}, M., {Louboutin}, M., {Kukreja}, N., {H{\"u}ckelheim},
  J., {Yount}, C., {Witte}, P., {Kelly}, P.~H.~J., {Gorman}, G.~J., and
  {Herrmann}, F.~J.: Architecture and performance of Devito, a system for
  automated stencil computation, CoRR, abs/1807.03032,
  \urlprefix\url{http://arxiv.org/abs/1807.03032}, 2018.

\bibitem[{McCalpin(1991-2007)}]{McCalpin2007}
McCalpin, J.~D.: STREAM: Sustainable Memory Bandwidth in High Performance
  Computers, Tech. rep., University of Virginia, Charlottesville, Virginia,
  \urlprefix\url{http://www.cs.virginia.edu/stream/}, a continually updated
  technical report. http://www.cs.virginia.edu/stream/, 1991-2007.

\bibitem[{McMechan(1983)}]{McMechan}
McMechan, G.~A.: MIGRATION BY EXTRAPOLATION OF TIME-DEPENDENT BOUNDARY VALUES,
  Geophysical Prospecting, 31, 413--420,
  \doi{10.1111/j.1365-2478.1983.tb01060.x},
  \urlprefix\url{http://dx.doi.org/10.1111/j.1365-2478.1983.tb01060.x}, 1983.

\bibitem[{Membarth et~al.(2012)Membarth, Hannig, Teich, and
  K{\"o}stler}]{HiPACC}
Membarth, R., Hannig, F., Teich, J., and K{\"o}stler, H.: Towards
  domain-specific computing for stencil codes in HPC, in: High Performance
  Computing, Networking, Storage and Analysis (SCC), 2012 SC Companion:, pp.
  1133--1138, IEEE, 2012.

\bibitem[{Meurer et~al.(2017)Meurer, Smith, Paprocki, \v{C}ert\'{i}k,
  Kirpichev, Rocklin, Kumar, Ivanov, Moore, Singh, Rathnayake, Vig, Granger,
  Muller, Bonazzi, Gupta, Vats, Johansson, Pedregosa, Curry, Terrel,
  Rou\v{c}ka, Saboo, Fernando, Kulal, Cimrman, and Scopatz}]{Sympy}
Meurer, A., Smith, C.~P., Paprocki, M., \v{C}ert\'{i}k, O., Kirpichev, S.~B.,
  Rocklin, M., Kumar, A., Ivanov, S., Moore, J.~K., Singh, S., Rathnayake, T.,
  Vig, S., Granger, B.~E., Muller, R.~P., Bonazzi, F., Gupta, H., Vats, S.,
  Johansson, F., Pedregosa, F., Curry, M.~J., Terrel, A.~R., Rou\v{c}ka, v.,
  Saboo, A., Fernando, I., Kulal, S., Cimrman, R., and Scopatz, A.: SymPy:
  symbolic computing in Python, PeerJ Computer Science, 3, e103,
  \doi{10.7717/peerj-cs.103},
  \urlprefix\url{https://doi.org/10.7717/peerj-cs.103}, 2017.

\bibitem[{Mittet(1994)}]{Mittet}
Mittet, R.: Implementation of the Kirchhoff integral for elastic waves in
  staggered-grid modeling schemes, GEOPHYSICS, 59, 1894--1901,
  \doi{10.1190/1.1443576}, \urlprefix\url{http://dx.doi.org/10.1190/1.1443576},
  1994.

\bibitem[{Naghizadeh and Sacchi(2009)}]{doi:10.1190/1.3008547}
Naghizadeh, M. and Sacchi, M.~D.: f-x adaptive seismic-trace interpolation,
  GEOPHYSICS, 74, V9--V16, \doi{10.1190/1.3008547},
  \urlprefix\url{https://doi.org/10.1190/1.3008547}, 2009.

\bibitem[{Osuna et~al.(2014)Osuna, Fuhrer, Gysi, and Bianco}]{osuna2014stella}
Osuna, C., Fuhrer, O., Gysi, T., and Bianco, M.: STELLA: A domain-specific
  language for stencil methods on structured grids, in: Poster Presentation at
  the Platform for Advanced Scientific Computing (PASC) Conference, Zurich,
  Switzerland, 2014.

\bibitem[{Patterson and Hennessy(2007)}]{Patterson}
Patterson, D.~A. and Hennessy, J.~L.: Computer Organization and Design: The
  Hardware/Software Interface, Morgan Kaufmann Publishers Inc., San Francisco,
  CA, USA, 3rd edn., 2007.

\bibitem[{Peters and Herrmann(2017)}]{peters2016cvp}
Peters, B. and Herrmann, F.~J.: Constraints versus penalties for
  edge-preserving full-waveform inversion, The Leading Edge, 36, 94--100,
  \doi{10.1190/tle36010094.1},
  \urlprefix\url{https://www.slim.eos.ubc.ca/Publications/Public/Journals/TheLeadingEdge/2016/peters2016cvp/peters2016cvp.html},
  (The Leading Edge), 2017.

\bibitem[{Plessix(2006)}]{PlessixASFWI}
Plessix, R.-E.: A review of the adjoint-state method for computing the gradient
  of a functional with geophysical applications, Geophysical Journal
  International, 167, 495--503, \doi{10.1111/j.1365-246X.2006.02978.x},
  \urlprefix\url{http://dx.doi.org/10.1111/j.1365-246X.2006.02978.x}, 2006.

\bibitem[{Raknes and Weibull(2016)}]{RaknesR45}
Raknes, E.~B. and Weibull, W.: Efficient 3D elastic full-waveform inversion
  using wavefield reconstruction methods, Geophysics, 81, R45--R55,
  \doi{10.1190/geo2015-0185.1},
  \urlprefix\url{http://geophysics.geoscienceworld.org/content/81/2/R45}, 2016.

\bibitem[{Rathgeber et~al.(2015)Rathgeber, Ham, Mitchell, Lange, Luporini,
  McRae, Bercea, Markall, and Kelly}]{Firedrake}
Rathgeber, F., Ham, D.~A., Mitchell, L., Lange, M., Luporini, F., McRae, A.
  T.~T., Bercea, G., Markall, G.~R., and Kelly, P. H.~J.: Firedrake: automating
  the finite element method by composing abstractions, CoRR, abs/1501.01809,
  \urlprefix\url{http://arxiv.org/abs/1501.01809}, 2015.

\bibitem[{Schmidt et~al.(2009)Schmidt, van~den Berg, Friedlander, and
  Murphy}]{SchmidtBergFriedlanderMurphy:2009}
Schmidt, M., van~den Berg, E., Friedlander, M.~P., and Murphy, K.: Optimizing
  Costly Functions with Simple Constraints: A Limited-Memory Projected
  Quasi-Newton Algorithm, in: Proceedings of The Twelfth International
  Conference on Artificial Intelligence and Statistics (AISTATS) 2009, edited
  by van Dyk, D. and Welling, M., vol.~5, pp. 456--463, Clearwater Beach,
  Florida, 2009.

\bibitem[{Seongjai~Kim(2007)}]{Kimapnum}
Seongjai~Kim, H.~L.: High-order schemes for acoustic waveform simulation,
  Applied Numerical Mathematics, 57, 402--414,
  \urlprefix\url{http://www.hl107.math.msstate.edu/pdfs/rein/HighANM_final.pdf},
  2007.

\bibitem[{Sun and Symes(2010)}]{sun2010iwave}
Sun, D. and Symes, W.~W.: IWAVE implementation of adjoint state method, Tech.
  rep., Tech. Rep. 10-06, Department of Computational and Applied Mathematics,
  Rice University, Houston, Texas, USA,
  \urlprefix\url{https://pdfs.semanticscholar.org/6c17/cfe41b76f6b745c435891ea6ba6f4e2c2dbf.pdf},
  2010.

\bibitem[{Symes(2015{\natexlab{a}})}]{symes2015acoustic}
Symes, W.~W.: Acoustic Staggered Grid Modeling in IWAVE, THE RICE INVERSION
  PROJECT, p. 141,
  \urlprefix\url{http://www.trip.caam.rice.edu/reports/2014/book.pdf#page=145},
  2015{\natexlab{a}}.

\bibitem[{Symes(2015{\natexlab{b}})}]{symes2015iwave}
Symes, W.~W.: IWAVE structure and basic use cases, THE RICE INVERSION PROJECT,
  p.~85,
  \urlprefix\url{http://www.trip.caam.rice.edu/reports/2014/book.pdf#page=89},
  2015{\natexlab{b}}.

\bibitem[{Symes et~al.(2011)Symes, Sun, and Enriquez}]{GPR:GPR977}
Symes, W.~W., Sun, D., and Enriquez, M.: From modelling to inversion: designing
  a well-adapted simulator, Geophysical Prospecting, 59, 814--833,
  \doi{10.1111/j.1365-2478.2011.00977.x},
  \urlprefix\url{http://dx.doi.org/10.1111/j.1365-2478.2011.00977.x}, 2011.

\bibitem[{Tang et~al.(2011)Tang, Chowdhury, Kuszmaul, Luk, and
  Leiserson}]{tang2011pochoir}
Tang, Y., Chowdhury, R.~A., Kuszmaul, B.~C., Luk, C.-K., and Leiserson, C.~E.:
  The pochoir stencil compiler, in: Proceedings of the twenty-third annual ACM
  symposium on Parallelism in algorithms and architectures, pp. 117--128, ACM,
  2011.

\bibitem[{Tarantola(1984)}]{Tarantola}
Tarantola, A.: Inversion of seismic reflection data in the acoustic
  approximation, GEOPHYSICS, 49, 1259, \doi{10.1190/1.1441754},
  \urlprefix\url{+ http://dx.doi.org/10.1190/1.1441754}, 1984.

\bibitem[{Umetani(1985)}]{deqsol}
Umetani, Y.: DEQSOL A numerical Simulation Language for Vector/Parallel
  Processors, Proc. IFIP TC2/WG22, 1985, 5, 147--164, 1985.

\bibitem[{Unat et~al.(2011)Unat, Cai, and Baden}]{unat2011mint}
Unat, D., Cai, X., and Baden, S.~B.: Mint: realizing CUDA performance in 3D
  stencil methods with annotated C, in: Proceedings of the international
  conference on Supercomputing, pp. 214--224, ACM, 2011.

\bibitem[{Van~Engelen et~al.(1996)Van~Engelen, Wolters, and Cats}]{ctadel}
Van~Engelen, R., Wolters, L., and Cats, G.: Ctadel: A generator of
  multi-platform high performance codes for pde-based scientific applications,
  in: Proceedings of the 10th international conference on Supercomputing, pp.
  86--93, ACM, 1996.

\bibitem[{van Leeuwen and Herrmann(2015)}]{vanleeuwen2015IPpmp}
van Leeuwen, T. and Herrmann, F.~J.: A penalty method for {PDE}-constrained
  optimization in inverse problems, Inverse Problems, 32, 015\,007,
  \urlprefix\url{https://www.slim.eos.ubc.ca/Publications/Public/Journals/InverseProblems/2015/vanleeuwen2015IPpmp/vanleeuwen2015IPpmp.pdf},
  (Inverse Problems), 2015.

\bibitem[{Versteeg(1994)}]{Versteeg927}
Versteeg, R.: The Marmousi experience; velocity model determination on a
  synthetic complex data set, The Leading Edge, 13, 927--936,
  \urlprefix\url{http://tle.geoscienceworld.org/content/13/9/927}, 1994.

\bibitem[{Virieux(1986)}]{Virieux86}
Virieux, J.: P-SV wave propagation in heterogeneous media: Velocity-stress
  finite-difference method, GEOPHYSICS, 51, 889--901, \doi{10.1190/1.1442147},
  \urlprefix\url{https://doi.org/10.1190/1.1442147}, 1986.

\bibitem[{Virieux and Operto(2009)}]{Virieux}
Virieux, J. and Operto, S.: An overview of full-waveform inversion in
  exploration geophysics, GEOPHYSICS, 74, WCC1--WCC26, \doi{10.1190/1.3238367},
  \urlprefix\url{http://library.seg.org/doi/abs/10.1190/1.3238367}, 2009.

\bibitem[{Wang et~al.(2017)Wang, Nissen, and Kreiss}]{Wang1090158}
Wang, S., Nissen, A., and Kreiss, G.: Convergence of finite difference methods
  for the wave equation in two space dimensions, Computing Research Repository,
  \urlprefix\url{https://arxiv.org/abs/1702.01383}, 2017.

\bibitem[{Warner and Guasch(2014)}]{AWI}
Warner, M. and Guasch, L.: Adaptive waveform inversion: Theory, pp. 1089--1093,
  \doi{10.1190/segam2014-0371.1},
  \urlprefix\url{https://library.seg.org/doi/abs/10.1190/segam2014-0371.1},
  2014.

\bibitem[{Wason et~al.(2017)Wason, Oghenekohwo, and
  Herrmann}]{wason2016GEOPctl}
Wason, H., Oghenekohwo, F., and Herrmann, F.~J.: Low-cost time-lapse seismic
  with distributed compressive sensing{\textendash}-{Part} 2: impact on
  repeatability, Geophysics, 82, P15--P30, \doi{10.1190/geo2016-0252.1},
  \urlprefix\url{https://www.slim.eos.ubc.ca/Publications/Public/Journals/Geophysics/2017/wason2016GEOPctl/wason2016GEOPctl.html},
  (Geophysics), 2017.

\bibitem[{Watanabe(2015)}]{Watanabe2015}
Watanabe, K.: Green's Functions for Laplace and Wave Equations, pp. 33--76,
  Springer International Publishing, Cham, \doi{10.1007/978-3-319-17455-6_2},
  \urlprefix\url{https://doi.org/10.1007/978-3-319-17455-6_2}, 2015.

\bibitem[{Weiss and Shragge(2013)}]{Weiss2013}
Weiss, R.~M. and Shragge, J.: Solving {3D} anisotropic elastic wave equations
  on parallel {GPU} devices., Geophysics, 78,
  \doi{http://dx.doi.org/10.1190/geo2012-0063.1}, 2013.

\bibitem[{Williams et~al.(2009)Williams, Waterman, and
  Patterson}]{williams2009roofline}
Williams, S., Waterman, A., and Patterson, D.: The Roofline model offers
  insight on how to improve the performance of software and hardware.,
  communications of the acm, 52, 2009.

\bibitem[{Witte et~al.(2018{\natexlab{a}})Witte, Louboutin, Lensink, Lange,
  Kukreja, Luporini, Gorman, and Herrmann}]{witte2018fwip3}
Witte, P., Louboutin, M., Lensink, K., Lange, M., Kukreja, N., Luporini, F.,
  Gorman, G., and Herrmann, F.~J.: Full-waveform inversion, Part 3:
  Optimization, \doi{10.1190/tle37020142.1},
  \urlprefix\url{https://doi.org/10.1190/tle37020142.1}, 2018{\natexlab{a}}.

\bibitem[{Witte et~al.(2018{\natexlab{b}})Witte, Louboutin, Kukreja, Luporini,
  , Lange, Gorman, and Herrmann}]{witte2018alf}
Witte, P.~A., Louboutin, M., Kukreja, N., Luporini, F., , Lange, M., Gorman,
  G.~J., and Herrmann, F.~J.: A large-scale framework for symbolic
  implementations of seismic inversion algorithms in Julia,
  \urlprefix\url{https://www.slim.eos.ubc.ca/Publications/Private/Submitted/2018/witte2018alf/witte2018alf.html},
  submitted to Geophysics on March 1, 2018., 2018{\natexlab{b}}.

\bibitem[{{Witte} et~al.(2018){Witte}, {Louboutin}, {Lange}, {Kukreja},
  {Luporini}, {Gorman}, and {Herrmann}}]{Witte2018b}
{Witte}, P.~A., {Louboutin}, M., {Lange}, M., {Kukreja}, N., {Luporini}, F.,
  {Gorman}, G., and {Herrmann}, F.~J.: A large-scale framework for symbolic
  implementations of seismic inversion algorithms in Julia, 2018.

\bibitem[{Yount(2015)}]{YASK}
Yount, C.: Vector Folding: Improving Stencil Performance via Multi-dimensional
  SIMD-vector Representation, in: 2015 IEEE 17th International Conference on
  High Performance Computing and Communications, 2015 IEEE 7th International
  Symposium on Cyberspace Safety and Security, and 2015 IEEE 12th International
  Conference on Embedded Software and Systems, pp. 865--870,
  \doi{10.1109/HPCC-CSS-ICESS.2015.27}, 2015.

\bibitem[{Zhang and Mueller(2012)}]{Autotuning-1}
Zhang, Y. and Mueller, F.: Auto-generation and Auto-tuning of 3D Stencil Codes
  on GPU Clusters, in: Proceedings of the Tenth International Symposium on Code
  Generation and Optimization, CGO '12, pp. 155--164, ACM, New York, NY, USA,
  \doi{10.1145/2259016.2259037},
  \urlprefix\url{http://doi.acm.org/10.1145/2259016.2259037}, 2012.

\end{thebibliography}

\end{document}